\documentclass[aps, prd, 10pt, twocolumn, longbibliography, amsfonts, amsmath, showkeys, nofootinbib, superscriptaddress, floatfix]{revtex4-2}

\usepackage[english]{babel}
\usepackage[utf8]{inputenc}
\usepackage{amsthm}
\usepackage{mathtools, bm}
\usepackage{physics}
\usepackage{xcolor}
\usepackage{graphicx} 
\usepackage{adjustbox}
\usepackage{placeins}
\usepackage[T1]{fontenc}
\usepackage{lipsum}
\usepackage{csquotes}
\usepackage[caption=false]{subfig}

\usepackage{tabularx}
\usepackage{hhline}
\usepackage{afterpage} 
\usepackage{array, booktabs, makecell}
\usepackage{multirow, bigdelim}
\usepackage{colortbl}
\usepackage{xspace}
\usepackage[normalem]{ulem}
\usepackage{comment}
\usepackage[pdftex, pdftitle={Article}, pdfauthor={Author}]{hyperref} 

\def\Msolar{M\ensuremath{_\odot}\xspace}
\newcommand{\asharp}{A$^\sharp$\xspace}
\newcommand{\aplus}{A+\xspace}

\usepackage[acronym]{glossaries}
\newcolumntype{P}[1]{>{\centering\arraybackslash}p{#1}}

\newacronym{ce}{CE}{Cosmic Explorer}
\newacronym{et}{ET}{Einstein Telescope}
\newacronym{cehs}{CEHS}{Cosmic Explorer Horizon Study}
\newacronym{nsf}{NSF}{National Science Foundation}
\newacronym{xg}{XG}{Next Generation}

\begin{document}
\title{A case study of GW190425 for classifying binary neutron star versus binary black hole mergers and constraining asymmetric dark matter with gravitational wave detectors}
\author{Sanika Khadkikar}
\affiliation{Institute for Gravitation and the Cosmos, Department of Physics, Pennsylvania State University, University Park, PA 16802, USA}
\author{Divya Singh}
\affiliation{Department of Physics, University of California, Berkeley, CA 94720, USA}

\begin{abstract}
The LIGO Scientific, Virgo, and KAGRA collaboration has identified two binary neutron star merger candidates, GW170817 and GW190425, along with several binary black hole candidates. While GW170817 was confirmed as a BNS merger through its electromagnetic counterparts, GW190425 lacked such observations, leaving its classification uncertain. We examine the possibility that GW190425 originated from black holes that merged after dark matter accretion caused their progenitor neutron stars to implode. Using this event, we place constraints on dark matter parameters, such as its mass and interaction cross section. We simulate GW190425-like events and analyze them using future gravitational wave detector networks, including upcoming upgrades to current detector networks and next-generation observatories. We show that a network with \aplus sensitivity can not classify a GW190425-like event with sufficient confidence. Detector networks with \asharp sensitivity can classify such events only if the neutron stars follow a relatively stiff equation of state, whose stronger tidal imprint differs measurably from a binary black hole waveform. Next-generation observatories like the Einstein Telescope and Cosmic Explorer recover the tidal signature even for soft, compact stars, enabling confident classification. Finally, we forecast the dark matter constraints that future gravitational wave networks could achieve for similar events.
\end{abstract}
\maketitle
\section{Introduction}
Gravitational wave (GW) detections have significantly advanced our understanding of the fundamental properties of compact objects, such as neutron stars (NSs) and black holes (BHs). So far, the LIGO Scientific, Virgo, and KAGRA (LVK) collaboration has detected many GW events with source parameters that are consistent with binary black hole (BBH), binary neutron star (BNS), and neutron star black hole (NSBH) mergers \cite{LIGOScientific:2018mvr, LIGOScientific:2020ibl, KAGRA:2021vkt}. The classification of GWs depends on several factors, such as the masses of the components, the presence of an electromagnetic (EM) counterpart, and the presence or absence of tidal information. The association of an EM counterpart along with the recovery of tidal information implies that at least one of the components in the binary is an NS. For example, GW170817 \cite{LIGOScientific:2017vwq, LIGOScientific:2018hze, LIGOScientific:2018cki} had a network signal-to-noise ratio (SNR) of $\sim$33 and a sky localization of $\sim$28 $\text{deg}^2$, allowing astronomers to detect an associated kilonova AT2017gfo \cite{Coulter:2017wya, DES:2017kbs} and a fast gamma-ray burst GRB170817A \cite{LIGOScientific:2017zic, Goldstein:2017mmi, Savchenko:2017ffs}. Together, the inferred tidal information and the multi-messenger constraints on GW170817 make it consistent with BNS data \cite{LIGOScientific:2017vwq}. However, there still exists a possibility that GW170817 was an NSBH merger \cite{LIGOScientific:2018hze, Hinderer:2018pei, Coughlin:2019kqf}. 

In contrast, GW190425 \cite{LIGOScientific:2020aai} had a lower network SNR of $\sim$13 and a much larger sky area of $\sim$8300 $\text{deg}^2$ as it was detected in only one of the detectors. Since it was farther away and poorly localized, an effective EM follow-up was not feasible. In such a case where no EM counterpart is available, classification relies entirely on information encoded in the GW signal. Among these, tidal deformability provides the most direct evidence for the presence of NSs, since such effects are unique to BNS systems and absent in BBH mergers~\cite{Damour:2009vw,Binnington:2009bb}. However, due to the low SNR of GW190425, strong constraints have not been placed on its possible tidal deformabilities, and thus it does not contribute significantly to its classification. Bayes factors could be used to compare the BBH and BNS hypotheses; however, the absence of a strong tidal signature to support the BNS hypothesis, combined with the fewer parameters required to describe the BBH model, makes the BBH hypothesis statistically more favorable.

In the absence of informative tidal constraints, existing knowledge about NS mass limits can still aid with classification. Theoretical calculations indicate that NSs have maximum masses $\lesssim$ 3 \Msolar \cite{Rhoades:1974fn, Kalogera:1996ci}. At the same time, X-Ray binaries suggest a lack of BHs observed having masses $\lesssim$ 5 \Msolar, indicating a "lower mass gap" in the $\sim$ 3 \Msolar to $\sim$ 5 \Msolar range \cite{Bailyn:1997xt, Ozel:2010su, Farr:2010tu}. However, recent observations of compact objects in non-interacting binaries \cite{Thompson:2018ycv, Jayasinghe:2021uqb}, an isolated compact object detected with microlensing \cite{Lam:2022vuq} and GW observations (GW230529 \cite{LIGOScientific:2024elc}, GW200115 \cite{LIGOScientific:2021qlt}, GW190814\cite{LIGOScientific:2020zkf}) indicate that the lower mass gap likely existed because of an observational bias. Using these prescribed ranges, compact objects having masses $\lesssim$ 3 \Msolar are generally consistent with NSs and others with BHs. However, \citet{Singh:2022wvw} show that accreting as little as one hundredth of a solar mass of asymmetric bosonic dark matter(DM) as massive as the WIMP, NSs can implode to form BHs. The rate of such implosions is influenced by the local DM density ($\rho_{\chi}$), interaction cross section of the DM particle with NS baryons ($\sigma_{\chi}$) and the mass of the DM particles ($m_{\chi}$). This suggests a novel formation channel for BHs with masses overlapping the previously identified NS mass range.

Since the component masses in GW190425 are largely consistent with $\lesssim 3 \, M_{\odot}$, it is widely considered a BNS merger candidate. However, incorporating the formation channel described in \citet{Singh:2022wvw} assigns a finite probability to GW190425 being an NSBH or BBH merger. We focus on GW190425 because its total mass ($\sim 3.4~M_\odot$) and chirp mass ($1.44~M_\odot$) lie about $5\sigma$ beyond the Galactic BNS distribution, yet they remain within the nominal NS range where only a small amount of dark matter can implode each star into a BH. \citet{Singh:2022wvw} note that BBH formation through this channel is far more likely than NSBH formation, since both neutron stars typically share the same environment, capture dark matter at similar rates, and implode on comparable timescales. Given the broad tidal posteriors and the lack of an associated EM counterpart, we investigate the possibility of an event like GW190425 being a BBH merger in this work and, under this assumption, place constraints on DM properties, specifically $m_\chi$ and $\sigma_\chi$.

Additionally, to test the robustness of this assumption and assess whether future detectors could enable confident classification, we simulate a GW190425-like event with upcoming detector networks. Planned upgrades of the current GW detectors to the \aplus \cite{T1800042} and \asharp \cite{T2200287} sensitivity and the advent of next-generation (XG) GW detectors like the Cosmic Explorer (CE) \cite{Reitze:2019iox, Evans:2021gyd, LIGOScientific:2016wof} and the Einstein Telescope (ET) \cite{Punturo:2010zza, Hild:2010id, ET:2019dnz} will operate at an increased sensitivity owing to their larger bandwidth and will also significantly improve the SNR. As a result, the recovery of tidal information. Due to their increased sensitivity, the SNR of a GW190425-like event in these future GW detector networks will be significantly higher. This enhanced sensitivity is expected to yield more reliable classification between BNS and BBH systems.

To quantify this, we simulate two sets of GW190425-like events using median parameters from the high-spin posterior distributions released by LVK \cite{LIGOScientific:2022, LIGOScientific:2019lzm, KAGRA:2023pio}. One set uses a BBH waveform model (\texttt{IMRPhenomXAS}~\cite{Pratten:2020fqn}), and the other uses a BNS waveform with tidal effects (\texttt{IMRPhenomXAS\_NRTidalv3}~\cite{Pratten:2020fqn, Abac:2023ujg}). Each set is analyzed across detector sensitivities using both waveform families. We then compute Savage–Dickey ratios \cite{Dickey1971} from the recovered tidal posteriors to assess classification accuracy.

Finally, assuming a future GW190425-like event is confidently identified as a BBH, we forecast the resulting improvement in DM constraints. These results demonstrate how future detectors will enhance both our ability to classify low-mass mergers and to use them as probes of DM interactions.

The rest of the paper is organized as follows. In Sec. \ref{sec2}, we provide a brief overview of the methods used in this study. Sec. \ref{sec2A} details the inference techniques, Sec. \ref{sec2B} discusses the interpretation of GW190425 as a BBH merger, and Sec. \ref{sec2C} outlines the detector networks and simulation settings employed to forecast DM constraints. In Sec. \ref{sec3}, we highlight the methodology for estimating and inferring constraints on DM parameters. Specifically, Sec. \ref{sec3A} estimates the observable merger rate, Sec. \ref{sec3B} describes the simulated population, and Sec. \ref{sec3C} briefly summarizes the collapse mechanism. The results of our analysis are presented in Sec. \ref{sec4}, and Sec. \ref{sec5} concludes with a summary and discussion of future directions.
\section{GW Inference Framework}
\label{sec2}
\subsection{Inferring the tidal information}
\label{sec2A}
The equation of state (EoS) characterizes the relationship between pressure and energy density inside a NS and plays a central role in determining its structure and composition. Depending on the choice of EoS, NSs can differ significantly in size and response to external forces. \textit{Soft} EoS models tend to produce smaller NSs that resist deformation, whereas \textit{stiff} EoS models result in larger stars that are more easily deformed. During a BNS merger, each NS deforms due to the tidal forces of its companion's gravitational field. The degree of this tidal deformation is quantified by a parameter known as the tidal deformability \cite{Hinderer:2007mb, Flanagan:2007ix}:
\begin{equation}\label{eq:lambda}
    \Lambda_i = \frac{2}{3}{k_2}_i(\frac{c^2R_i}{Gm_i})^5
\end{equation}

Here, ${k_2}_i$ denotes the tidal Love number, $m_i$ is the mass, and $R_i$ is the radius of the $i^{\text{th}}$ NS. The constants $c$ and $G$ represent the speed of light and the gravitational constant, respectively. The dimensionless tidal deformability $\Lambda$ is strictly positive for NSs and expected to be zero for BHs, making it a useful parameter for distinguishing between BNS and BBH mergers.

GW signals from BBH mergers have typically been modeled using 11 source parameters within the quasi-circular orbit approximation to infer the properties of the source. These include intrinsic parameters such as masses $m_i$ and spin components aligned with orbital angular momentum $\chi_i$. Furthermore, sky location $(\alpha, \delta)$, polarization angle $\psi$, inclination angle $i$, luminosity distance $D_L$, coalescence time $t_c$ and phase $\phi_c$ are the seven extrinsic parameters that are also inferred. For a BNS GW signal, the two tidal deformabilities $\Lambda_i$ are also included in the model. We then infer posterior probability densities for all of these parameters that form the set $\vec{\theta}$.

In the frequency domain, the measured strain in response to a GW signal from a coalescing system can be expressed as
\begin{equation}\label{eq:wf}
    \tilde{h}(f) = \mathcal{A}(f)e^{i\psi(f)}
\end{equation}
where $\mathcal{A}(f)$ is the GW amplitude and $\psi(f)$ is the GW phase~\cite{PhysRevD.44.3819}. Tidal deformation during the inspiral leads to additional energy dissipation, modifying the GW phase evolution at the fifth post-Newtonian (5PN) order~\cite{Flanagan:2007ix,Favata:2013rwa}. 
\begin{equation}\label{eq:5PN}
    \psi_{\rm{Tidal}}(f) = -\frac{39}{2} \tilde{\Lambda}v^{10} + \left( \frac{6595}{364}\delta \tilde{\Lambda} - \frac{3115}{64}\tilde{\Lambda} \right) v^{12}
\end{equation}
Here, $v=(\pi M f)^{1/3}$ is the expansion co-efficient (assuming c=1) expressed in terms of the total mass of the binary, $M \equiv m_1 + m_2$ and the GW frequency, $f$.
This leading order tidal term depends not on the component tidal deformabilities of the NSs in the binary, but on the mass ratio weighted sum of the individual tidal deformabilities as shown below
\begin{equation}\label{eq:lambda_tilde}
    \tilde{\Lambda} = \frac{16}{13} \frac{(12q +1)\Lambda_1 + (12 + q)q^4\Lambda_2}{(1+q)^5}
\end{equation}
Here $q=m_2/m_1$ is the mass ratio with $m_1$ being the mass of the heavier NS. This tidal contribution becomes significant in the final few cycles of the BNS coalescence and can be measured if the GW signal is detected with a sufficiently high SNR.

\begin{figure}[t!]
    \centering
    \includegraphics[width=\linewidth]{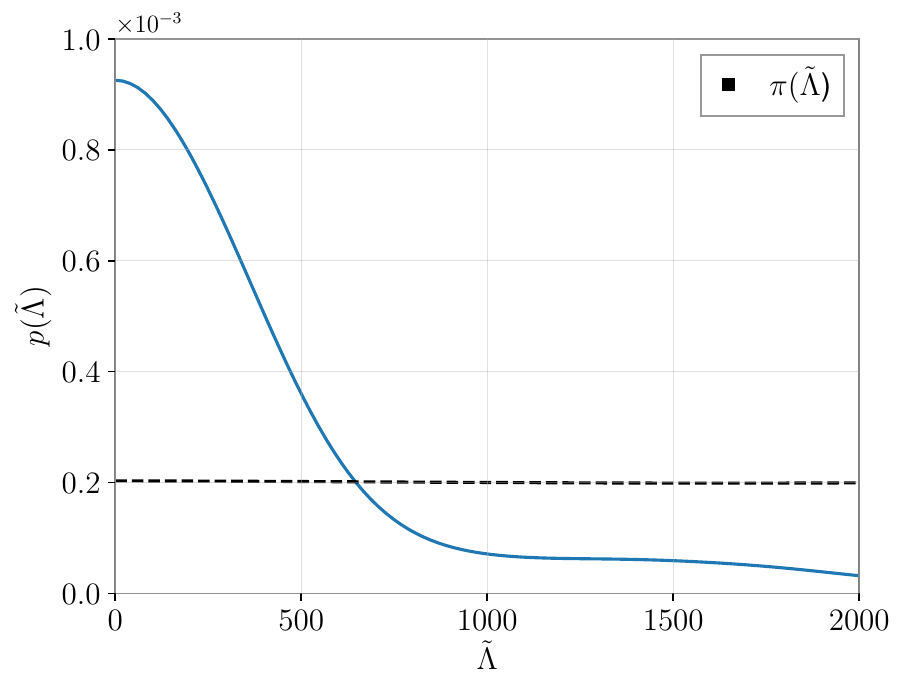}
    \caption{Posterior distribution of the tidal deformability parameter $\tilde{\Lambda}$ for GW190425, inferred using data from the GWTC-2.1 catalog \protect \cite{LIGOScientific:2021usb}, is plotted in blue. The prior distribution is shown as a black dotted line. The posterior peaks sharply near $\tilde{\Lambda} = 0$ and falls off toward larger values. This behavior suggests that GW190425 may have originated from a BBH merger.}
    \label{fig:gw190425_lambda}
\end{figure}

We use the Bayesian inference library $\tt{bilby}$ \cite{Romero-Shaw:2020owr, Ashton:2018jfp, lalsuite, swiglal} and the $\tt{dynesty}$ \cite{Speagle:2019ivv} sampler to estimate the parameters $\vec{\theta}$ from observed GW data. According to Bayes' theorem, the posterior probability distribution of the parameters $\vec{\theta}$ in the data $d$ is given as
$$p(\vec{\theta}|d) = \frac{\mathcal{L}(d|\vec{\theta})\pi(\vec{\theta})}{Z_{GW}}$$
where $\mathcal{L}(d|\vec{\theta})$ is the likelihood function, $\pi(\vec{\theta})$ are the prior probability distributions of the parameters $\vec{\theta}$ and $Z_{GW}$ is the model evidence. We use $\tt{bilby}$ to obtain $p(\vec{\theta}|d)$ and employ its relative binning implementation to speed up likelihood evaluations in our analysis \cite{Cornish:2010kf, Cornish:2021lje, Zackay:2018qdy}.

\subsection{GW190425 as a Binary Black Hole Merger}
\label{sec2B}
As discussed earlier, tidal interactions provide a key observable to distinguish NS mergers from BBH systems, since tidal effects are absent in BBHs. To investigate this in the context of GW190425, we analyze publicly available posterior samples from the GWTC-2.1 catalog \cite{LIGOScientific:2021usb}.

Figure \ref{fig:gw190425_lambda} shows the posterior distribution for the effective tidal deformability, $\tilde{\Lambda}$, inferred from GW190425. The distribution peaks at $\tilde{\Lambda} = 0$, consistent with the absence of tidal effects expected in a BBH merger. While the relatively low SNR of this event limits the precision with which tidal effects can be constrained, the posterior still favors values near zero. This behavior, when combined with the absence of a detected EM and the inferred component masses falling outside the typical Galactic NS mass distribution \cite{Antoniadis:2016hxz, Alsing:2017bbc, 2020RNAAS...4...65F, Shao:2020bzt, Landry:2021hvl}, strengthens the possibility that GW190425 may have been a BBH merger. Thus, in this study, we proceed under the assumption that the source of GW190425 was a BBH merger to derive constraints on DM properties.

\begin{figure}[!b]
    \centering
    \includegraphics[width=\linewidth]{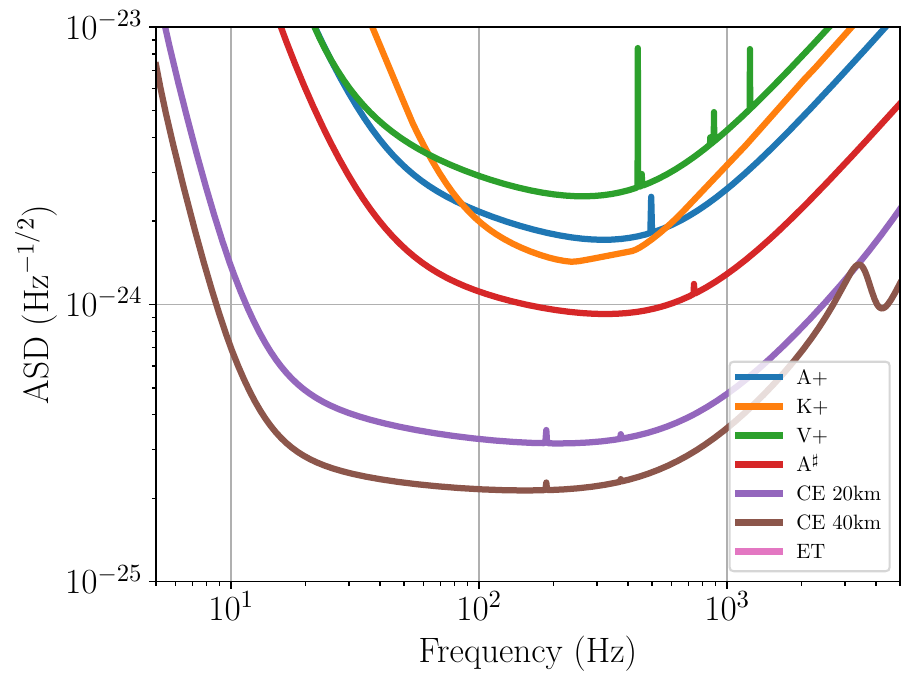}
    \caption{Amplitude spectral density (ASD) of detector noise for LIGO (A+ \cite{T1800042} and A$^\sharp$ \cite{T2200287} sensitivity), Virgo (V+), KAGRA (K+), Cosmic Explorer (20 km and 40 km), and Einstein Telescope (ET) \cite{ET-0304B-22}.} 
    \label{fig:asds}
\end{figure}

\subsection{Forecasting Future GW190425-like events}
\label{sec2C}
GW190425 requires the assumption that it is a BBH due to the limited SNR of the event, which prevents a definitive measurement of $\tilde{\Lambda}$. However, with planned upgrades to current detectors and the advent of XG gravitational wave observatories, future GW190425-like events are expected to be observed with significantly higher SNRs. This will enable for a more precise inference of $\tilde{\Lambda}$ and allow for a more definitive classification of the source. Thus, we also provide forecasts using the projected sensitivities of several detector sensitivities as follows.

\subsubsection{Detector Networks}\label{subsub:nets}
In this study, we consider three different detector networks to analyze simulated signals following the inferred source properties of GW190425 using Bayesian inference. Additionally, to infer the collapse time from the observation of a GW190425-like signal, we also perform extensive simulations using the Fisher formalism (Sec. \ref{sec3B}) to estimate errors on measured properties which inform the observable merger rate in the corresponding detector networks. These results are further cast into DM constraints.
The corresponding noise spectra are presented in Fig. \ref{fig:asds}.

\FloatBarrier
\begin{itemize}
    \item \textbf{HLVK+}: The projected sensitivity of the current LVK detectors for the fifth observation run \cite{KAGRA:2013rdx}. This network includes LIGO Hanford, LIGO Livingston~\cite{LIGOScientific:2014pky}, KAGRA~\cite{10.1093/ptep/ptaa125}, and Virgo~\cite{VIRGO:2014yos} detectors with A+ sensitivity. 
    \item \textbf{HLI\#}: The next upgrade of the LIGO detectors after A+ referred to as A$^\sharp$\cite{T2200287}. For this configuration, we have also considered LIGO-India in addition to LIGO Hanford and LIGO Livingston at A$^\sharp$ sensitivity. LIGO-India is a new GW interferometer currently being constructed in Aundha, India, and it is initially expected to achieve \aplus sensitivity. However, we assume an upgraded LIGO India operating with $A^{\sharp}$ sensitivity \cite{Saleem:2021iwi}.
    \item \textbf{CE4020ET}: To analyze the most optimistic case, we use XG observatories including one triangular ET \cite{Punturo:2010zza, Hild:2010id, ET:2019dnz}, one CE 20 km observatory, and one CE 40 km observatory \cite{Reitze:2019iox, Evans:2021gyd, LIGOScientific:2016wof}. We position the 40 km CE observatory in Hanford and the 20 km observatory in Livingston at the exact locations of the current LIGO interferometers. ET is positioned in Cascina in Italy, at the current location of the Virgo interferometer. 
\end{itemize}

\subsubsection{Simulation Setup}
We generate both BNS and BBH signals like the event GW190425, to assess their distinguishability using future upgrades and generations of GW detectors. To simulate BNS signals, we use the \texttt{IMRPhenomXAS\_NRTidalv3}~\cite{Pratten:2020fqn, Abac:2023ujg} waveform model, which augments the closed-form, phenomenological, quadrupolar \texttt{IMRPhenomXAS}~\cite{Pratten:2020fqn} model by incorporating tidal effects relevant for quasi-circular BNS inspirals. For BBH signals, we employ the \texttt{IMRPhenomXAS} model itself, which includes inspiral, merger, and ringdown. The intrinsic and extrinsic parameters for these simulations are chosen to be the median values obtained from the posterior distributions in the GWTC-2.1-confident data release by the LVK collaboration~\cite{ligo_scientific_collaboration_and_virgo_2022_6513631}. The injected signal parameters as summarized in Table~\ref{tab:injection_values}.
\begin{table}[htb]
\caption{Source parameter values chosen from GWTC-2.1 confident data release~\cite{ligo_scientific_collaboration_and_virgo_2022_6513631} used to simulate a GW190425-like gravitational wave signal using the $\tt{IMRPhenomXAS}$ and $\tt{IMRPhenomXAS\_NRTidalv3}$ waveform in \texttt{bilby}.}\label{tab:injection_values}
\begin{ruledtabular}
\begin{tabular}{lc}
    \textbf{Parameter} & \textbf{Injected Value}\\
    \hline
    Component masses $m_1$, $m_2$     &  2.08 \Msolar, 1.32 \Msolar\\
    Dimensionless spins $\chi_1$, $\chi_2$    &  0.07, 0.04 \\
    Luminosity Distance $D_L$     & 160 Mpc \\
    Right ascension $\alpha$ & 3.16  \\
    Declination $\delta$ & $-0.07$ \\
    Inclination $\iota$ & 1.70 \\
    Polarization $\psi$ & 0.57 \\
\end{tabular}
\end{ruledtabular}
\end{table}

For the least sensitive detector network considered in this study, HLVK+, simulating the signal using the median sky location parameters from the GWTC-2.1 posterior placed the source in a region of low network sensitivity, resulting in an SNR  $\sim 5$ which is below the detection threshold. Given that HLVK+ is expected to be more sensitive than the network that originally detected GW190425, this outcome is not likely for a GW190425-like event. To ensure that the injected signal is detectable and provides a meaningful comparison, we use sky location parameters corresponding to the maximum likelihood sample from the posteriors for the HLVK+ simulation. This choice reflects an optimistic scenario in which the HLVK+ network observes a GW190425-like signal.

\begin{figure}[htb]
    \includegraphics[width=1.0\linewidth]{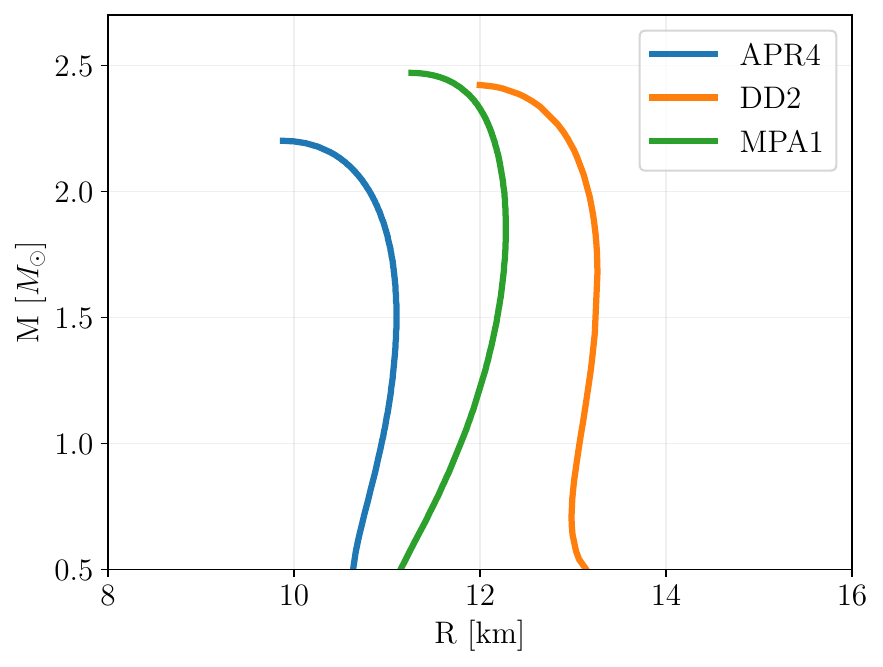}
    \caption{Mass–radius curves for the APR4, MPA1, and DD2 EoSs used to simulate future GW190425-like events under the assumption that the source was a BNS merger. APR4, the softest EoS considered, yields highly compact NSs with low tidal deformability $\tilde\Lambda$, resulting in a BNS system that closely mimics a BBH in its gravitational-wave signature. In contrast, DD2 is a relatively stiff EoS that produces less compact stars with larger $\tilde\Lambda$, thereby enhancing the distinguishability of the system from a BBH merger.}
    \label{fig:eoss}
\end{figure}

To compute $\Lambda_i$ associated with $m_i$ in Table~\ref{tab:injection_values} for the BNS simulations, we use three representative EoS: APR4 \cite{Akmal:1998cf}, MPA1 \cite{Muther:1987xaa}, and DD2 \cite{Typel:2009sy, Hempel:2009mc}. These EoSs span a broad range in $m$–$R$ space and are consistent with constraints from GW170817 \cite{LIGOScientific:2018cki}. APR4 is the softest among them, DD2 the stiffest, and MPA1 represents an intermediate case, as illustrated in Fig.~\ref{fig:eoss}.
We have adopted low-spin priors ($\chi \leq 0.05$) for all BNS simulations and broader high-spin priors ($\chi \leq 0.89$) for BBH simulations \cite{LIGOScientific:2018hze, LIGOScientific:2020aai}. The priors on the component masses are uniform, and the tidal deformability parameters $\Lambda_1$ and $\Lambda_2$ are also assigned uniform priors, bounded between [0, 5000] and [0, 10000], respectively. The tidal parameter $\Lambda_2$ corresponding to the secondary mass $m_2$ has a broader range owing to the inverse relationship between them \cite{LIGOScientific:2018cki}. All other priors follow the default settings in \texttt{bilby} which are detailed in Table \ref{tab:priors}. We perform parameter estimation using the \texttt{dynesty} sampler, with the number of walks set to 100 and the maximum number of MCMC steps set to 5000. We use 2048 live points and set the number of accepted steps per chain(naccept) to 60.
\begin{table}[htb]
\caption{Priors used in the Bayesian estimation of signal parameters to study BNS signals.}
\label{tab:priors}
\begin{ruledtabular}
\begin{tabular}{lc}
    \textbf{Parameter} & \textbf{Distribution} \\
    \hline
    $\cos \theta_{\rm JN}$ & Uniform in [-1, 1]  \\
    Polarization $\psi$ & Uniform in [0, $\pi$] \\
    Right ascension $\alpha$ & Uniform in \\
    &[0, $2\pi$] \\
    Declination $\delta$ & Cosinusoidal in \\
    &[$-\pi/2, \pi/2$] \\
    Luminosity Distance $D_L$ & Uniform\ in Comoving\ Volume \\
    & in [0, 1000] Mpc \\
\end{tabular}
\end{ruledtabular}
\end{table}

\section{Dark Matter Parameter Inference Framework}
\label{sec3}
\subsection{Observable Merger Rate}
\label{sec3A}
For the population of BBHs that form through implosion due to DM accumulation in the cores of NSs, we estimate the observable BBH merger rate for the detector networks listed in Sec. \ref{subsub:nets}.
This observable rate, $N_{\rm BBH}$, is a function of:
\begin{enumerate}
    \item the uncertainty in the measurement of the effective tidal deformability signified by ($\sigma^{90\%}_{\tilde{\Lambda}}$), and
    \item the collapse time for the NS to become a BH ($t_c$)
\end{enumerate}
The measured 90\% confidence interval of the effective tidal deformability, $\sigma_{\tilde{\Lambda}}^{90\%}$, is used as the discriminator between a BNS and BBH signal. Therefore, it is used in conjunction with the SNR to build the detector network efficiency,
\begin{equation}
    \epsilon(z) = \frac{1}{N} \sum_{i=1}^N \Pi \left( \frac{\rm{SNR}}{\rm{SNR}_T} - 1 \bigg | z \right) \Pi \left( \frac{\sigma_{\tilde{\Lambda}_T}}{\sigma_{\tilde{\Lambda}}} - 1 \bigg | z \right)
\end{equation}
which encodes the capability of a detector network to not only detect a GW signal assuming a threshold SNR, $\rm{SNR_T}$ but also distinguish whether the source is a BBH or BNS for an assumed threshold $\sigma_{\tilde{\Lambda}_T}$.

We use the Fisher formalism~\cite{Cutler:1994ys,Poisson:1995ef} to derive the expected SNR and $\sigma_{\tilde{\Lambda}}$ for a population of BBH, which is described in detail in Sec. \ref{sec3B}.
Using the Fisher informed efficiency, we compute the expected number of BBH mergers for the considered networks, as a function of $t_c$ and fiducial thresholds on $\sigma_{\tilde{\Lambda}}^{90\%}$.
\begin{equation}\label{eq:NBBH}
    N_{\rm BBH} = \dot{N}_{\rm BBH} T = T \int_0^z \frac{\dot{n}(z')_{\rm BBH}}{1 + z'} \frac{dV_c}{dz'} \epsilon(z') dz'
\end{equation}
Here $T$ is the observing period over which the detector network is expected to operate.
For this study, we set $T = 1$ year, unless otherwise noted.
The dependence of the merger rate on $t_c$ comes through the merger rate density, $\dot{n}(z)_{\rm BBH}$, which is derived by integrating the star-formation rate over all possible delay times weighted by a delay time distribution, $\mathcal{P}(t_d) \propto t_d^{-1}$.
\begin{equation}\label{eq:ndotBBH}
    \dot{n}(z)_{\rm BBH} = A \int_{t_c}^{t_d^{\rm max}} \psi(z_f(z,t_d))\mathcal{P}(t_d)dt_d
\end{equation}
Here, delay time refers to the time between the formation of NS to the merger of the BNS system.
The delay times of interest for this BBH population span over $[t_c, t_d^{\rm max}]$ since the NSs in the binary must implode to form BHs before the maximum time of merger, $t_d^{\rm max}$ in order to contribute to $N_{\rm BBH}$.
Following~\citet{Singh:2022wvw}, we restrict $t_d$ within 20 million years and the Hubble time.
In Eq. \ref{eq:ndotBBH}, the constant $A$ is set such that the merger rate density of all binary systems with component masses $\leq 3~M_\odot$, $\dot{n}(0)$ is equal to the inferred BNS merger rate of $105.5^{+190.2}_{-83.9}~\rm{per~Gpc^3~per~yr}$ using data from GWTC-3~\cite{KAGRA:2021duu}.
We assume the star formation rate, $\psi(z)$ from Ref.~\cite{Madau:2014bja}.

From the perspective of GW searches, any compact object binary with components that have masses $\leq 3~M_\odot$ is classified as a BNS.
However, if the detected signal is in fact from a BBH formed through the implosion scenario, using a tidal waveform like \texttt{IMRPhenomXAS\_NRTidalv3} for a BBH signal can produce a non-zero inference for the effective tidal deformability parameter as discussed in Sec. ~\ref{sec4A}.
In our method, we use those constraints as our threshold on $\sigma_{\tilde{\Lambda}}$, when determining the collapse time from the observed number of BNS-like signals with $\tilde{\Lambda} \simeq 0$.

Note that the inference of the collapse time is DM-model agnostic.
Within some model, the $t_c$ is a function of $\rho_\chi$, $m_\chi$ and $\sigma_\chi$.
Therefore, if we can infer $t_c$ from GW observations, we can constrain the parameter space of DM properties within the assumed model.

\subsection{Fisher Analysis}
\label{sec3B}

To forecast the relation between the collapse time, $N_{\rm{BBH}}$, $t_c$ and $\sigma_{\tilde{\Lambda}}$, we first use the Fisher matrix formalism to empirically derive the distribution of errors in the measurement of source parameters for the desired population of BBHs.
Although Bayesian inference is the preferred method for inferring uncertainties, the Fisher approach is fast and computationally efficient for the large parameter space and number of detector networks that we explore here.
Since the formation mechanism follows the implosion of NS with as little as a hundredth of a solar mass of DM accumulated at their cores, the considered BBH population closely resembles the population of BNSs considered widely in literature~\cite{Gupta:2023lga}.
Table~\ref{tab:fisher_pop} lists the range of source parameters forming the source population. The redshift distribution of sources follows the star formation rate from Ref.~\cite{Madau:2014bja} with local merger rate density of $105.5^{+190.2}_{-83.9}~\rm{per~Gpc^3~per~yr}$~\cite{KAGRA:2021duu}.
\begin{table}[htb]
\caption{Source parameters of the BBH population used in the Fisher analysis.}
\label{tab:fisher_pop}
\begin{ruledtabular}
\begin{tabular}{lc}
    \textbf{Parameter} & \textbf{Source Distribution} \\
    \hline
    Component masses, $m_i$   & Double gaussian within \\
    & [1, 2.2] \Msolar \\
    Component spins, $a_i$    &  Uniform in [0, 0.1] \\
    Right ascension, $\alpha$ & $[0, 2\pi]$ \\
    Declination, $\delta$ & $[-\pi/2 , \pi/2 ]$ \\
    Inclination, $\iota$ & $[0, \pi]$ \\
    Polarization, $\psi$ & $[0, 2\pi]$ \\
\end{tabular}
\end{ruledtabular}
\end{table}

The Fisher approximation starts with computing the Fisher information matrix $\Gamma_{mn} = \langle \partial_m \tilde{h}, \partial_n \tilde{h} \rangle $.
Here, $\partial_m \tilde{h}$ is the partial derivative of the Fourier transform of the detector response(Eq.~\ref{eq:wf}) with respect to parameter $\lambda_m$.
Each matrix element is a scalar product $\langle a, b \rangle$ defined as
\begin{equation}
    \langle a, b \rangle \equiv 4 \mathfrak{R} \int df \frac{a(f) b^{*}(f)}{S_n(f)}
\end{equation}
where $S_h(f)$ is the one-sided power spectral density characteristic to the considered detector(Fig. \ref{fig:asds}).

The information matrix for a detector network is computed by summing over the individual detector matrices.
Finally, the covariance matrix is derived by inverting the network information matrix $C_{mn} = \Gamma^{-1}_{mn}$.
The diagonal elements of the covariance matrix, $C_{mm}$ encode the variances of the parameters, $\sigma^2_{\lambda_m}$, while the off-diagonal elements are the covariances between different parameters. 
For more details on the use of Fisher formalism in GW data analysis, the reader is referred to Ref.~\cite{Cutler:1994ys,Poisson:1995ef}.

For this study, we only focus on the SNR and $\sigma_{\tilde{\Lambda}}^{90^\%}$ for each source.
We compute $\sigma_{\tilde{\Lambda}}^{90\%}$ for the population described in Table~\ref{tab:fisher_pop} using the \texttt{gwbench} toolkit~\cite{Borhanian:2020ypi}. We compute these errors for the detector network that observed GW190425, hereafter labeled O3, and for the networks described in Sec. \ref{subsub:nets}.
It can be noted from Eq.~\ref{eq:5PN} that $\sigma_{\tilde{\Lambda}}^{90\%}$ derived from the Fisher information matrix is independent of the exact value of $\tilde{\Lambda}$ as it appears linearly in its leading post-Newtonian term.
This is not true for Bayesian inference where we find that smaller tides due to a soft EoS can produce larger uncertainties in the measurement of $\tilde{\Lambda}$~\cite{Khadkikar:2025ith}.
We neglect the sub-dominant effects of dynamical tides from the second tidal term (sixth PN correction) \cite{Lai:1993di, Steinhoff:2016rfi}.

\begin{figure*}[!t]
    \centering
    \includegraphics[width=\linewidth]{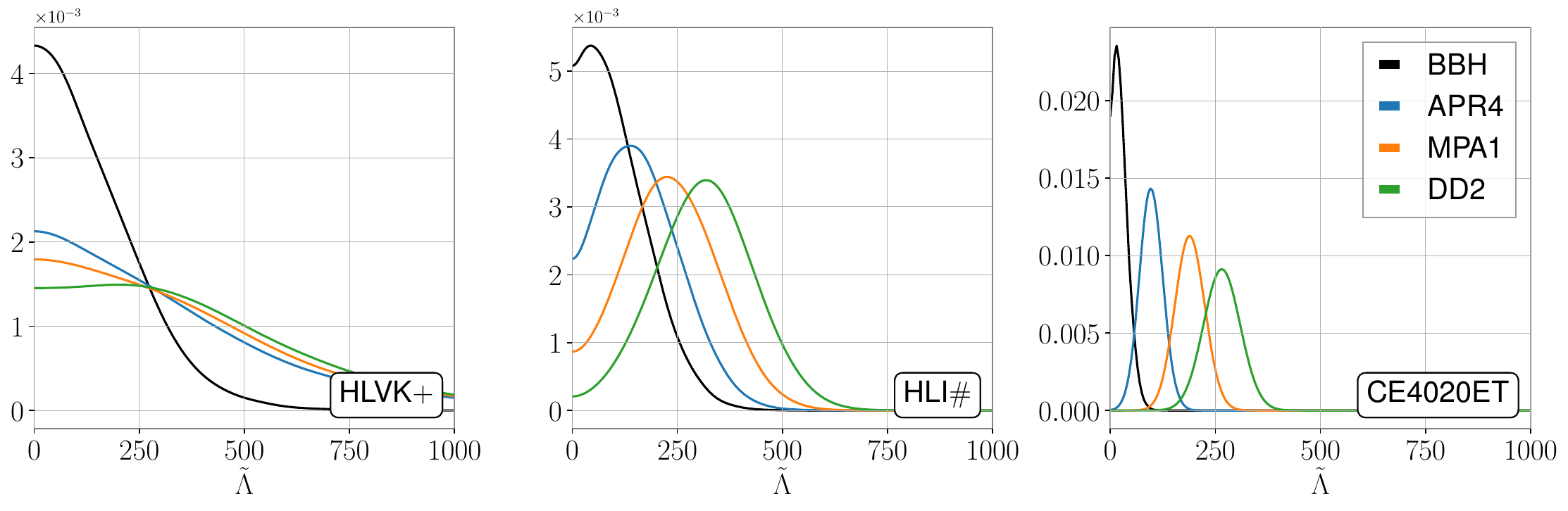}
    \caption{Forecasted posterior distributions of the tidal deformability parameter $\tilde{\Lambda}$ using simulated GW190425-like signals across detector networks of increasing sensitivity (left to right). Colored curves correspond to simulations based on different BNS equations of state: APR4 (blue), MPA1 (orange), and DD2 (green), all analyzed with a BNS waveform model. The black curve represents a BBH simulation analyzed using a BNS waveform. While only one scenario occurs in reality, clear separation between the posteriors aids in distinguishing BNS and BBH sources. For the HLVK+ network, the posteriors exhibit limited separation. In contrast, the HLI$\#$ network allows reliable classification if NSs are known to favor a stiff EoS. The CE4020ET network provides the clearest separation, with all BNS-based posteriors showing negligible support at $\tilde{\Lambda} = 0$, offering the strongest capability for source classification.}
\end{figure*}
\subsection{Inferring dark matter properties}
\label{sec3C}
We consider the implosion mechanism detailed in ~\citet{Singh:2022wvw} to derive limits on the DM parameter space using the GW observation of a GW190425-like event in the considered detector networks.
The $t_c$ inferred from the observation of GW190425 and GW190425-like event(s) in different networks is used to constrain $m_\chi$ and $\sigma_\chi$ through the capture of DM in the NS core and the eventual implosion of the NS to a BH.
Following the mechanism detailed in ~\citet{Singh:2022wvw}, we consider the asymmetric DM sector where ambient bosonic DM particles stream through the NS and get eventually captured, and thermalize within its core.
This thermalized sphere self-gravitates and forms a BH if it exceeds the Chandrasekhar limit. 
The BH then consumes the host star, provided it survives Hawking radiation.
For $m_\chi \leq 2 \times 10^4$ GeV, the mini-BH forms through the Bose-Einstein condensation of DM. 
However, the critical mass of BH to survive Hawking radiation and continue to grow is only reached for $m_\chi \leq 10$ GeV.
In the scenario where the BH forms without Bose-Einstein condensation, $m_\chi > 2 \times 10^4$ GeV, the BH is more massive than the critical mass required to avoid evaporation, and the NS undergoes implosion.

The dominant timescale for BH formation in either formation channels is given by Eq.~\ref{eq:tc BEC} and~\ref{eq:tc accretion}.

\begin{equation}
\begin{split}
          t^{\rm BEC}_{\rm BH} \approx 5.2\times 10^4 \left(\frac{\rho_0}{\rho_\chi}\right) \left(\frac{\rm GeV}{m_\chi}\right) {\rm Max} \left[\frac{\sigma_{\rm th}}{\sigma},1 \right] \\
          \times \left[ 1+ \frac{2}{3} \left(\frac{m_\chi}{10 ~\rm GeV}\right)^2 \left(\frac{T}{10^5~{\rm K}}\right)^3 \right] {\rm yr}
\end{split}
\label{eq:tc BEC}
\end{equation}

\begin{equation}
\begin{split}
    t_{\rm BH} \approx 10^{13}~\left(\frac{\rho_0}{\rho_\chi}\right) \left(\frac{\rm GeV}{m_\chi}\right)^{3/2} {\rm Max} \left[\frac{\sigma_{\rm th}}{\sigma},1 \right]~{\rm yr}
\end{split}
\label{eq:tc accretion}
\end{equation}

Here, we use $\rho_0 = 1 \rm{GeV/cm^3}$ as the fiducial value for $\rho_\chi$; $\sigma_{\rm th}$ is the threshold value of the cross-section determined by the geometry of the NS such that the DM particles traversing the NS have at least one collision.
For a NS with mass of $1.4~M_\odot$ and radius 12 km and total baryon number $\simeq 2 \times 10^{57}$, $\sigma_{\rm th} \approx 2 \times 10^{-45} ~{\rm cm}^2$.
For more details, the reader is referred to Sec. VI in~\citet{Singh:2022wvw}.

\section{Results}\label{sec4}

In this section, we detail the results from the various parts of the inference framework.
We first look at the results from the Bayesian parameter estimation studies for GW190425-like events with the considered detector networks in Sec.~\ref{sec4A}.
Next we briefly discuss the the population-wide Fisher parameter estimation results with emphasis on the distribution of SNR and $\sigma^{90\%}_{\tilde{\Lambda}}$ of the sources in Sec.~\ref{sec4B}.
In Sec.~\ref{sec4C}, we present the observable merger rates for each network derived using the Fisher estimates, which are used to infer the $t_c$ from the observation of a GW190425-like BBH event and the posteriors for $\tilde{\Lambda}$ from Bayesian PE.
Lastly, we use the inferred $t_c$ to derive constraints on DM properties within the model considered for this formation channel in Sec.~\ref{sec4D}.

\subsection{PE Comparison\label{sec4A}}

In this section, we present the results of our parameter estimation runs for GW190425 and for simulated GW190425-like events observed with future detector networks. The aim is to assess whether tidal information, specifically the posterior distribution of $\tilde{\Lambda}$ and its uncertainty $\sigma_{\tilde{\Lambda}}$, can distinguish a GW signal from a BNS merger versus one originating from a system that may have undergone the implosion process to form a BBH system. We simulate signals using both BNS and BBH waveforms, analyze them with BNS waveform templates, and evaluate the resulting posteriors across three different detector network configurations described in Sec. \ref{subsub:nets}.

Bayes factors are generally used to test whether the GW source favors the BBH or BNS hypothesis. Since the BBH model has two fewer parameters than the BNS model, the Occam factor penalizes the BNS hypothesis and can tip the Bayes factor toward BBH even when the data favors non-zero tidal effects. To avoid this bias, we adopt the Savage-Dickey ratio (see Appendix~\ref{appendix:SD}) and the posterior width $\sigma_{\tilde{\Lambda}}^{90\%}$ as our classification metrics. A potential indicator is how strongly the posterior peaks at $\tilde{\Lambda}=0$ and the sharpness of that peak, both captured by these two quantities. The results are summarized in Table~\ref{tab:results}.
\begin{table*}
\centering\caption{Summary of signal properties inferred across detector networks. For each case, we report the SNR, Savage-Dickey ratio, and the measured effective tidal deformability $\tilde{\Lambda}$ for BBH and BNS signals generated with different EoSs.}
\label{tab:results}
\begin{ruledtabular}
\begin{tabular}{llccc}
\textbf{Detector Network} & \textbf{System Type} & \textbf{SNR} & \textbf{Savage-Dickey Ratio} & $\boldsymbol{\tilde{\Lambda}}$ \\
\midrule
\multirow{4}{*}{HLVK+} 
  & BBH & 23.01 & 21.65 & $120.3^{+238.7}_{-113.7}$ \\
  & BNS (APR4) & 22.85 & 10.63 & $263.6^{+661.2}_{-245.0}$ \\
  & BNS (MPA1) & 23.24 & 8.97 & $300.4^{+675.3}_{-277.5}$ \\
  & BNS (DD2) & 23.06 & 7.26 & $336.4^{+662.8}_{-307}$ \\
\midrule
\multirow{4}{*}{HLI$\#$} 
  & BBH & 33.72 & 25.38 & $90.9^{+153.1}_{-86.1}$ \\
  & BNS (APR4) & 33.45 & 11.18 & $149.7^{+176.0}_{-132.8}$ \\
  & BNS (MPA1) & 33.68 & 4.34 & $228.3^{+189.5}_{-181.2}$ \\
  & BNS (DD2) & 34.02 & 1.04 & $311.8^{+193.1}_{-199.1}$ \\
\midrule
\multirow{4}{*}{CE4020ET} 
  & BBH & 304.70 & 95.22 & $18.1^{+30.2}_{-18.1}$ \\
  & BNS (APR4) & 304.71 & 5.2 & $94.6^{+46.7}_{-45.3}$ \\
  & BNS (MPA1) & 304.66 & 2.42 $\times$ $10^{-5}$ & $186.8^{+59.1}_{-57.5}$ \\
  & BNS (DD2) & 304.19 & < $10^{-15}$ & $262.9^{+72.2}_{-71.6}$ \\
\end{tabular}
\end{ruledtabular}
\end{table*}

For GW190425, the Savage-Dickey ratio exceeds one, indicating some support for the posterior peaking at $\tilde{\Lambda} = 0$. However, the width $\sigma_{\tilde{\Lambda}}$ remains large, and the posterior is broad, as shown in Fig.~\ref{fig:gw190425_lambda}. This prevents confident classification and reflects the limited sensitivity of the detector network at the time. In the HLVK+ network, the SNR is similar to that of GW190425, leading to only marginal improvement in tidal constraints. Posteriors for both BNS and BBH simulations remain broad, and Savage-Dickey ratios remain similar across both cases. Despite additional detectors, the marginal difference in sensitivity limits classification.

The HLI$\#$ network yields a higher SNR, resulting in significantly tighter constraints on $\tilde{\Lambda}$. For soft EoSs like APR4, the BNS posterior continues to overlap with the BBH posterior, leaving the classification unresolved. For stiffer EoSs such as MPA1 and DD2, the differences are more pronounced. The BNS posterior shifts away from zero, becomes narrower, and clearly separates from the BBH posterior.

If future observations support a stiff NS EoS, the HLI$\#$ network can provide a useful classification framework. Under the BBH-merger hypothesis, the Savage–Dickey ratio is about 25 for our simulated BBH injection but about 1 for the simulated BNS injection that uses the DD2 EoS. These values correspond to $\sim 96\%$ probability of a BBH origin for the BBH injection and $\sim 51\%$  for the BNS injection (see Appendix \ref{appendix:SD}). These probabilities are not computed for earlier cases due to their broad posteriors.

The CE4020ET network delivers the most precise and accurate results. Even for APR4, the BNS posterior lies well away from zero and shows minimal overlap with the BBH case. For MPA1 and DD2, the classification becomes unambiguous. The posterior widths decrease significantly, and the support at $\tilde{\Lambda} = 0$ becomes negligible. In this setting, the Savage-Dickey ratio for the simulated BBH case yields a $\sim 99\%$ probability of it being a BBH signal. For the simulated BNS signal created with DD2, the probability of it being a BBH signal is close to zero; for MPA1, it is about $0.001\%$, and for APR4, around $19\%$. Although soft EoSs retain some ambiguity, the CE4020ET network enables robust classification of such edge cases. The main conclusion of this analysis is that if future observational or theoretical evidence supports stiffer EoSs, both HLI$\#$ and CE4020ET will allow classification of low-mass systems as either BBHs or BNSs. In the absence of strong EoS constraints, CE4020ET alone provides sufficient sensitivity for confident classification.

\subsection{Fisher Analysis}
\label{sec4B}

\begin{figure}[htb]
    \centering
    \includegraphics[width=\linewidth]{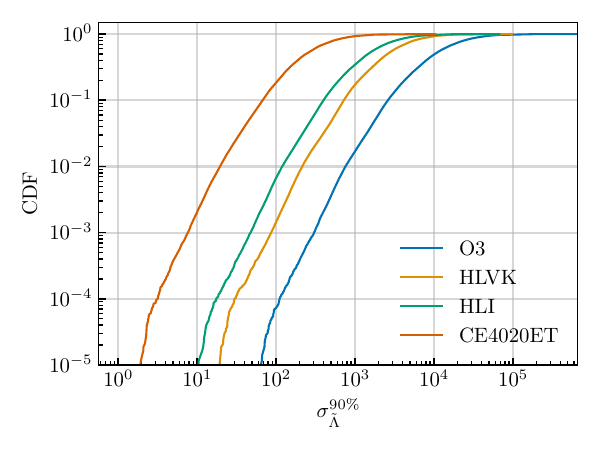}
    \caption{Cumulative distribution function of $\sigma^{90\%}_{\tilde{\Lambda}}$ in the detector networks considered in this work. Note that the fraction of events detected with an error $\sigma^{90\%}_{\tilde{\Lambda}} \simeq 100$ is 0.07\% for O3, 0.15\% for HLVK+, 0.65\% for HLI$\#$ and 20\% for CE4020ET respectively.}
    \label{fig:cdf_lamerr}
\end{figure}

For the purpose of this work, we are largely interested in the distribution of SNR and $\sigma^{90\%}_{\tilde{\Lambda}}$ for the BBH source population.
Results from the Fisher analysis consistently show that the detectability (SNR) and differentiability of BNS from BBH ($\sigma^{90\%}_{\tilde{\Lambda}}$) improves across the detector networks as the detectors get more sensitive, as observed in results from Bayesian inference(see Table~\ref{tab:results}) as well.
Fig. \ref{fig:cdf_lamerr} shows that with CE4020ET, $\tilde\Lambda$ can be measured to within $\pm 825$ (90$\%$ credible interval) for 90$\%$ of simulated sources.
For HLVK+, HLI$\#$ and O3 networks, 90\% of the simulated sources have $\sigma^{90\%}_{\tilde{\Lambda}}$ 4996, 8408 and 37210 respectively.
The fraction of events detected with a fiducial $\sigma^{90\%}_{\tilde{\Lambda}} \simeq 100$ is 0.07\% for O3, 0.15\% for HLVK+, 0.65\% for HLI$\#$ and 20\% for CE4020ET respectively.
We find that $\sigma^{90\%}_{\tilde{\Lambda}}$ can be as low as \(2\) for some high SNR events in the CE4020ET network. Thus, at this level of precision, the specific treatment of tidal parameters can introduce biases that equal or exceed the statistical uncertainty and noticeably shift the inferred $\tilde{\Lambda}$. This emphasizes the need for waveform models with much higher tidal fidelity to mitigate such biases once CE4020ET becomes operational \cite{Gamba:2020wgg}.

\subsection{Inferring Collapse Time}
\label{sec4C}

Following \citet{Singh:2022wvw}, we infer the collapse time from the observation of a potential BBH signal formed through this implosion channel using the Bayesian parameter estimate of $\tilde{\Lambda}$ reported in Table~\ref{tab:results}.
We first compute the detectable merger rates($N_{\rm BBH}$) as a function of fiducial $t_c$ and thresholds on SNR and $\sigma_{\tilde{\Lambda}}$.


\begin{figure*}[t]
  \centering
  \subfloat[]{
    \includegraphics[width=\linewidth]{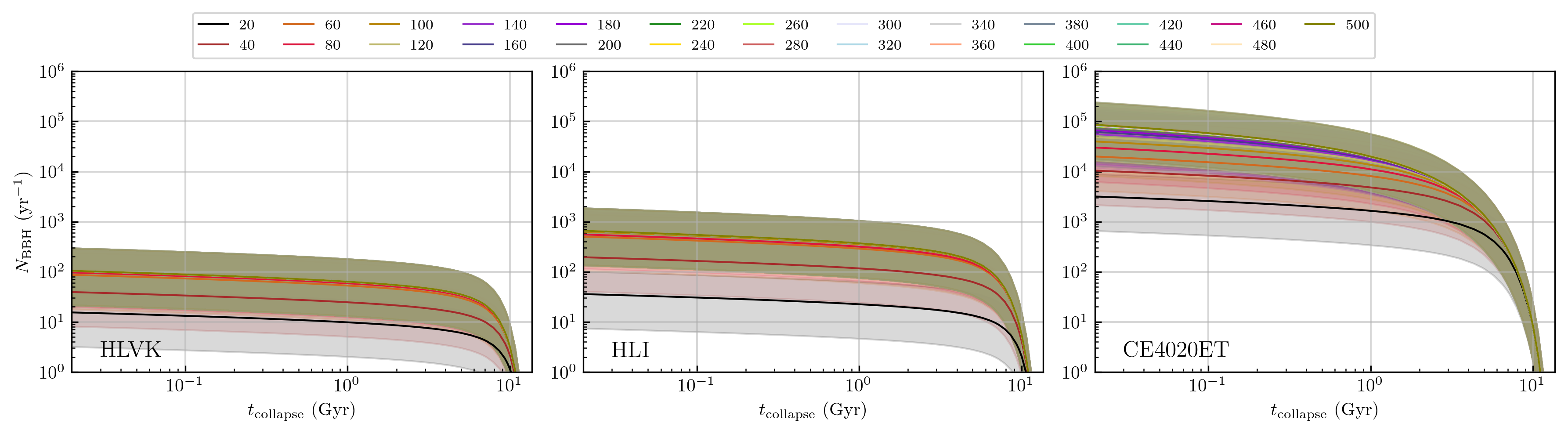}
    \label{fig:Ndot_snr8}
  }\par
  \subfloat[]{
    \includegraphics[width=\linewidth]{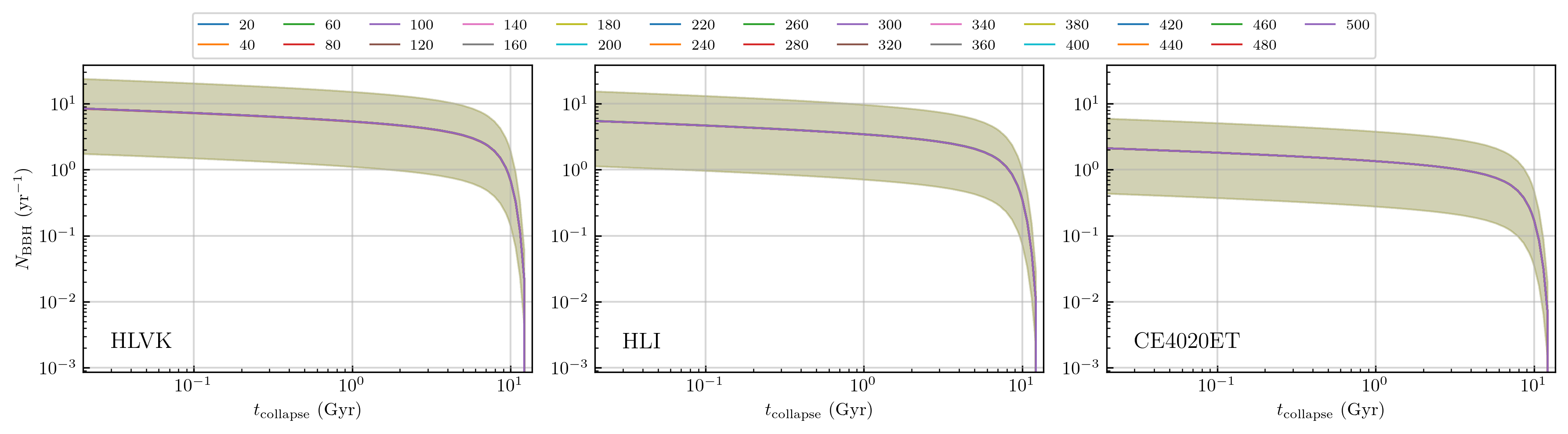}
    \label{fig:Ndot_snr-event}
  }
  \caption{Detectable merger rate of BBH mergers as a function of collapse time and the threshold on $\sigma^{90\%}_{\tilde{\Lambda}}$ (a) when $\rm{SNR_T} = 8$ for all networks, and (b) when $\rm{SNR_T}$ is equal to the SNR of the GW190425-like BBH signal in the respective detector networks(see Table~\ref{tab:results}).}
  \label{fig:merger_rates}
\end{figure*}

Figure~\ref{fig:Ndot_snr8} shows how the rates converge to lower values for networks with lower sensitivity due to lower efficiency, $\epsilon(z | \rm{SNR_T}, \sigma_{\tilde{\Lambda}_T})$, when the same SNR threshold($\rm{SNR_T} = 8$) is applied.
As the sensitivity of the GW detectors in the network increases, more sources are detectable with higher SNRs and lower $\sigma_{\tilde{\Lambda}}$.
Therefore, the overall observable rates are highest for the CE4020ET network at every $\sigma_{\tilde{\Lambda}_T}$ for a common SNR threshold($\rm{SNR_T} = 8$) across all networks.
The bands in Fig.~\ref{fig:Ndot_snr8} correspond to the uncertainty in the local merger rate of BNS mergers, $150.5^{+190.2}_{-83.9}~\rm{Gpc^{-3} yr^{-1}}$ from GWTC-3~\cite{KAGRA:2021duu}.

\begin{figure}
    \centering
    \includegraphics[width=\linewidth]{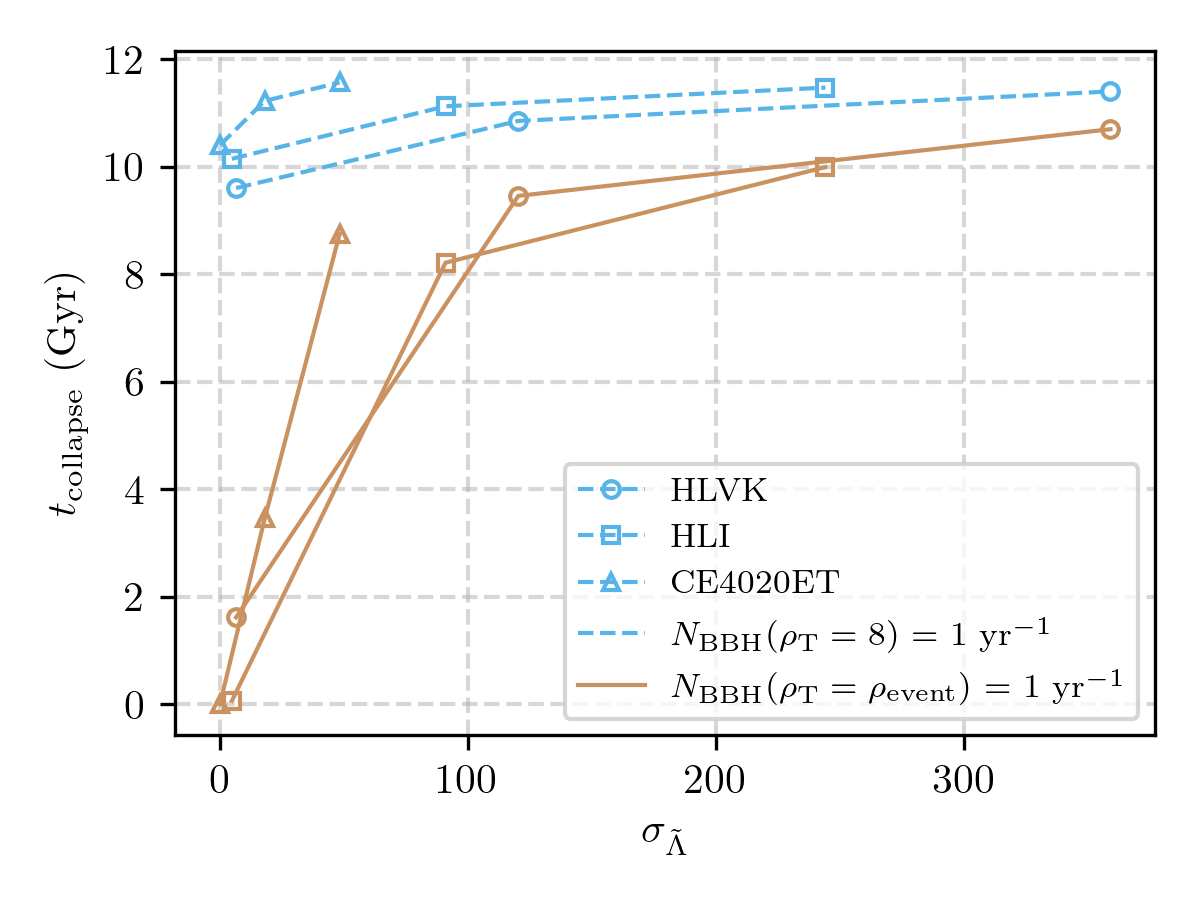}
    \caption{Inferred collapse time for $N_{\rm {BBH}}=1$ per yr for $\rm{SNR_T} =8$(\textit{blue}) and $N_{\rm {BBH}}=1$ for $\rm{SNR_T}$ set to event SNR in respective detector networks given in table~\ref{tab:results}(\textit{yellow}). For both cases, $\sigma^{90\%}_{\tilde{\Lambda}}$ is equal to the $95^{\rm th}$ percentile value for the GW190425-like BBH signal from table~\ref{tab:results} for the respective detector networks.}
    \label{fig:tc_err}
\end{figure}

Using the estimated $N_{\rm BBH}$, we infer the collapse times through an interpolated function such that $t_{\rm collapse} = t_{\rm collapse}(N_{\rm BBH}=1 | \rm{SNR_T} = 8, \sigma_{\tilde{\Lambda}_T})$ where $\sigma_{\tilde{\Lambda}_T}$ is informed by the Bayesian estimate of $\tilde{\Lambda}$ for the GW190425-like signal under the BBH hypothesis.
We interpret the $95^{\rm th}$ percentile of the Bayesian estimate to be the deviation in $\tilde{\Lambda}$ away from $\tilde{\Lambda}_{\rm BBH} = 0$.
We use $\sigma_{\tilde{\Lambda}_T}$ to infer a range of $t_{\rm collapse}$ for each detector network as shown in Fig.~\ref{fig:tc_err} where the range of $t_{\rm collapse}$ corresponds to the uncertainty band in $N_{\rm BBH}$ that arises from the uncertainty in the local BNS merger rate.
With an increase in network sensitivity, we find higher $t_c$ with less uncertainty if only one such BBH event is observed, which points towards a lower overall rate of such BBH mergers with less likelihood of this formation scenario.
$N_{\rm BBH}(\rm{SNR_T} = 8) = 1$ provides a conservative limit on the collapse time given that the observable rates account for every detectable source irrespective of its dissimilarity from the GW190425-like event. 
Note that this inference assumes an observing period of 1 year for all detector networks shown in Fig.~\ref{fig:Ndot_snr8}.

We also consider the case where the observable rates are conditioned by the event-statistics of the GW190425-like BBH event.
We use the SNR of this event as listed in Table~\ref{tab:results} to compute the efficiency, $\epsilon(z | \rm{SNR_T}, \sigma_{\tilde{\Lambda}_T})$ and the expected BBH observable rate.
We find much fewer signals in the Fisher population with event statistics similar to the GW190425-like BBH signal, especially for CE4020ET where this event has an SNR of 304.
Due to the stringent SNR threshold, $N_{\rm BBH}$ converges to the same value across $\sigma_{\tilde{\Lambda}_T}$ as seen in Fig.~\ref{fig:Ndot_snr-event}.
In this case, $N_{\rm BBH}$ is expectedly much lower than $N_{\rm BBH}(\rm{SNR_T} = 8)$.
However, we find that the expected rate is lowest for CE4020ET out of the considered networks given the unusually high SNR of this event, along with the higher precision and accuracy in the measurement of $\tilde{\Lambda}$.

Given the lower overall rates for GW190425-like events, a single event with a precise measurement of $\tilde{\Lambda}$ provides a lower estimate of $t_c$ implying that the NS implode to BH earlier in the merger.
Following the minimal BBH rate in CE4020ET, a singular detection gives the lowest range of inferred $t_c$ out of the considered networks.
The unertainty in $t_c$ arising from the uncertainty in the local BNS merger rate, is still highest for HLVK+ and lowest for CE4020ET, which directly follows from the increasing sensitivity across the detector networks.

\subsection{Dark Matter Constraints}
\label{sec4D}

We constrain the interaction cross-section of asymmetric bosonic DM with baryons within the NS, $\sigma_\chi$ for $m_\chi \in [1, 2e5]$ GeV using the inferred $t_c$ from the observation of GW190425 and GW190425-like BBH events in O3, HLVK+, HLI$\#$ and CE4020ET detector networks.
In doing so, we consider the time taken by the DM particles to get captured, thermalize and collapse to form a mini BH massive enough to avoid Hawking evaporation through Bond-Hoyle accretion and accretion of ambient DM.
Using Eqs.~\ref{eq:tc BEC} and~\ref{eq:tc accretion}, we produce limits on $\sigma_\chi$ for $m_\chi < 2 \times 10^4$ GeV considering the scenario where the mini BH forms through BEC formation, whereas for $m_\chi > 2 \times 10^4$ GeV, the DM mass becomes self-gravitating and collapses to a BH before the BEC temperature is achieved.

\begin{figure}[htb]
    \centering
    \includegraphics[width=\linewidth]{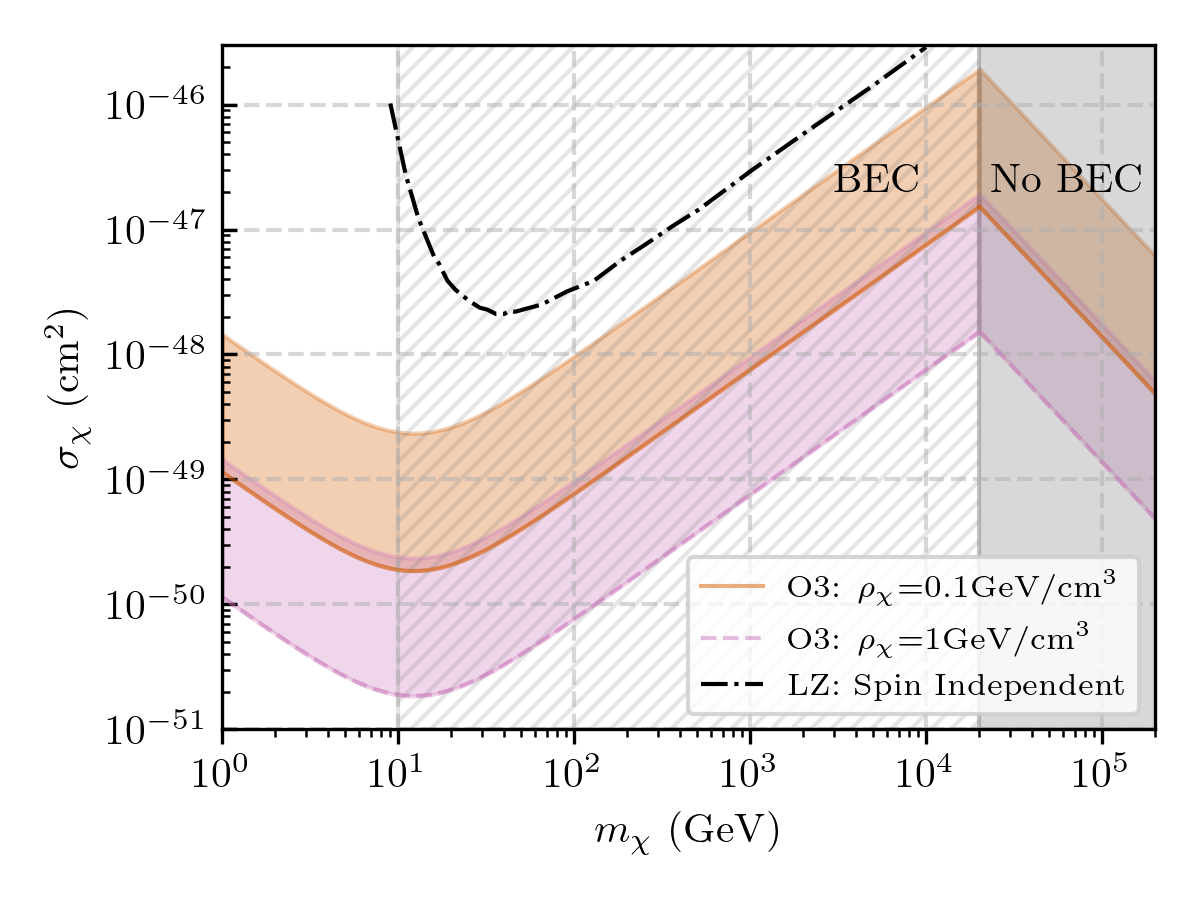}
    \caption{Limits on the scattering cross-section, $\sigma_\chi$ as a function of mass of DM particle, $m_\chi$ using the collapse time, $t_c$ inferred from the detection of GW190425 in O3. We use $\rm{SNR_T} =8$ and $\sigma_{\tilde{\Lambda}_T} = 1327$ to determine $N_{\rm {BBH}}(t_c)$ for the O3 observing period, $T=0.9$ yr~\cite{KAGRA:2021vkt}. We present limits for a fiducial ambient DM density, $\rho_\chi$ of 1 $\rm{GeV/cm^3}$ and $\rho_\chi = $ 0.1 $\rm{GeV/cm^3}$ which is of the order of $\rho_\chi$ near the Sun. The latest LZ WIMP-DM limits are plotted for reference\cite{LZ:2024zvo,hepdata.155182.v2}.}
    \label{fig:dm_limits_O3-event}
\end{figure}

\begin{figure}[ht]
  \centering
  \subfloat[]{%
    \includegraphics[width=\linewidth]{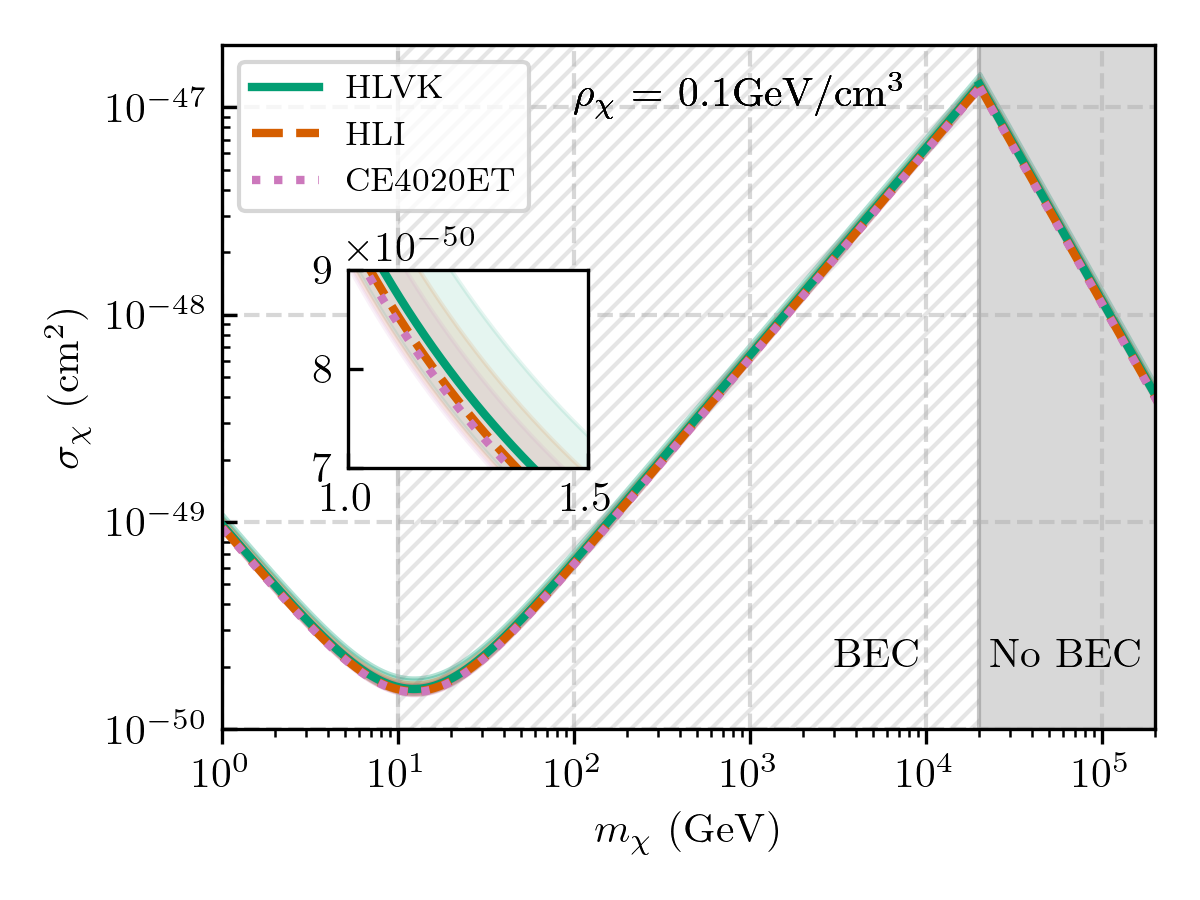}%
    \label{fig:dm_limits_all-8}%
  }\par

  \subfloat[]{%
    \includegraphics[width=\linewidth]{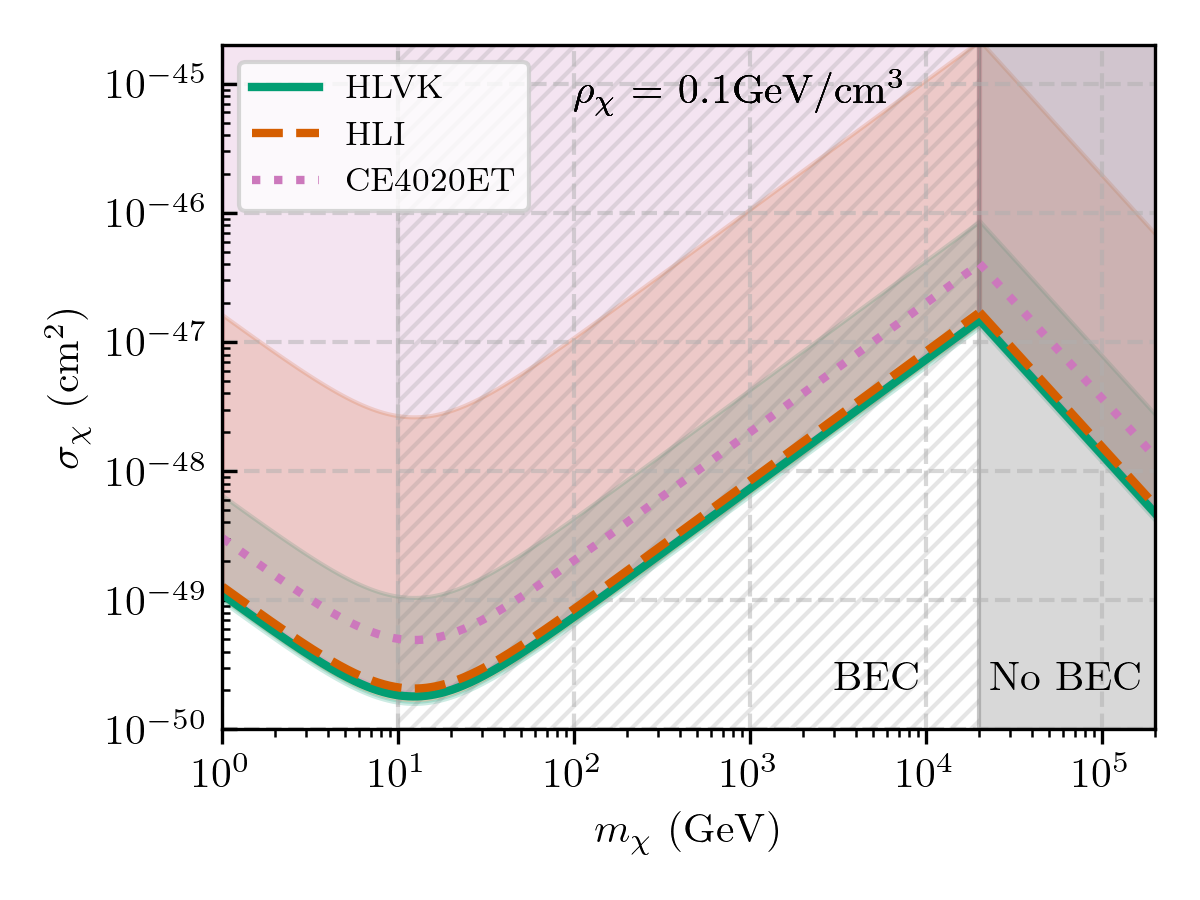}%
    \label{fig:dm_limits_all-event}%
  }

  \caption{Limits on the scattering cross section $\sigma_\chi$ as a function of the dark-matter particle mass $m_\chi$, using the collapse time $t_c$ inferred from GW190425. Panel (a) sets $\mathrm{SNR}_T=8$ for all networks. Panel (b) uses $\mathrm{SNR}_T$ equal to the SNR of the GW190425-like BBH signal in each detector network (see Table~\ref{tab:results}).}
\label{fig:dm_limits}
\end{figure}

Note that the critical mass of BH is only reached for $m_\chi < 10$ GeV through the BEC formation, therefore the limits shown in Figs.~\ref{fig:dm_limits_O3-event},~\ref{fig:dm_limits_all-8} and~\ref{fig:dm_limits_all-event} are valid outside of $m_\chi \in [10, 2 \times 10^4]$ GeV.

Figure~\ref{fig:dm_limits_O3-event} shows the limits derived from the observation of GW190425 in O3.
As evident in Fig.~\ref{fig:gw190425_lambda}, the posterior distribution for $\tilde{\Lambda}$ extends to $\sim 1000$, but does not exclude 0.
We use the $95^{\rm th}$ percentile from the posterior distribution of $\tilde{\Lambda}$ to compute the predicted $N_{\rm BBH}(t_c | \rm{SNR_T} = 8, \sigma_{\tilde{\Lambda}} = 1327)$ across fiducial $t_c$.
These predicted rates are then used to infer $t_c$ given $N_{\rm BBH} = 1$ over the O3 observing period of $\sim 0.9$ year~\cite{KAGRA:2021vkt}.
Along with these constraints, we show the latest limits on the WIMP cross-section from the direct detection experiment, LZ~\cite{LZ:2024zvo,hepdata.155182.v2}.
These limits lie within the range of $m_\chi$ where the BH evaporates in this NS implosion scenario.
However, we see how the astrophysical observation of imploding NS can be complementary to constraining the weak interaction of GeV-mass DM with visible matter.

In figures~\ref{fig:dm_limits_all-8} and~\ref{fig:dm_limits_all-event}, we project constraints for future detector configurations using $t_c$ inferred from the GW190425-like BBH signals.
These constraints follow from the inference of $t_c$ shown in Fig~\ref{fig:tc_err} for both cases.
For predicted rates obtained when $\rm{SNR_T} = 8$, a singular observation of GW190425-like BBH event produces the most stringent upper limits for CE4020ET.
The range of $\sigma_\chi$ at every $m_\chi$ directly follows from the uncertainty in inferred $t_c$ for all networks.
We find that CE4020ET provides the strictest upper limits on $\sigma_\chi$ if we observe only one BBH event forming through this implosion channel given the high observable predicted rates as shown in Fig~\ref{fig:Ndot_snr8}.
On the other hand, for predicted merger rates of GW190425-like BBH events i.e. when $\rm{SNR_T}$ is chosen to be the observed SNR of  GW190425-like BBH event in Table~\ref{tab:results}, we find the most spread in the upper limits for CE4020ET with the lower bound on the $\sigma_\chi$ being unconstrained due to the extremely low predicted rate for GW190425-like BBH event which can be interpreted as the NS merging before implosion($t_c <$ minimum delay time).
The most stringent upper limits are derived for HLVK+ following from the high inferred $t_c$.

To summarize the trends we observe, a singular detection of a GW190425-like BBH event when the predicted BBH merger rates are high results in a stringent limit on $\sigma_\chi$ over the range of $m_\chi$ considered here.
Since the $t_c$ inferred from this method is restricted by the age of the universe, the inferred upper limits on the interaction cross-section can only be explored down to $\sigma_\chi(t_{\rm Hubble}, \rho_\chi, m_\chi)$.
Given this restriction, we find that in either case, we derive the most information about DM properties from a single GW observation with the next generation of detectors, CE4020ET.


\section{Conclusions}
\label{sec5}

This work explores the hypothesis that GW190425 may have been a BBH merger formed through dark matter–induced collapse, and uses this scenario to place constraints on dark matter properties. Using publicly available LVK posterior samples for GW190425, we obtain limits on $m_\chi$ and $\sigma_\chi$ assuming various $\rho_\chi$ values. To assess the robustness of this assumption, we also simulate GW190425-like signals under both BBH and BNS scenarios and analyze them with appropriate waveform models across more sensitive future detector networks. This allows us to quantify the extent to which improved sensitivity enhances constraints on $\tilde{\Lambda}$ and, in turn, improves our ability to distinguish between BBH and BNS systems in the absence of an EM counterpart.

Our results show that, for current detector sensitivities and networks like HLVK+, the classification of GW190425 as a BNS or BBH remains uncertain due to the event’s low SNR and thus weak tidal signature. However, upgraded networks such as HLI$\#$ and XG networks like CE4020ET are forecasted to improve classification accuracy substantially. In particular, we find that HLI$\#$ could enable confident distinction between BBH and BNS scenarios, provided that NSs are known by then to favor a stiff EoS. If NSs instead follow a softer EoS description, the classification remains ambiguous with the HLI$\#$ network.

In contrast, the CE4020ET network offers robust classification for such cases. We find that the $\tilde{\Lambda}$ posterior for a BNS system simulated with even the softest EoS considered (APR4) remains distinguishable from the BBH case, with all BNS-based posteriors showing negligible support at $\tilde{\Lambda} = 0$. This highlights the potential of XG detectors to resolve the nature of ambiguous low-mass mergers using tidal information alone.

Assuming a confident detection of a GW190425-like event in future detectors, we forecast the resulting constraints on dark matter parameters. We examine two cases, each assuming a single low-mass BBH detection ($N_{\rm{BBH}} = 1$) but differing in the SNR threshold applied. The first adopts the standard LVK threshold of $\mathrm{SNR}_T = 8$, while the second sets the threshold equal to the SNR that GW190425-like BBH would have in the future detector network.

In the low-threshold scenario, we consider the case where signals with $\mathrm{SNR} > 8$ are detectable, but only a single event is observed. This suggests that the dark matter–induced collapse mechanism must be relatively slow, suppressing the formation of additional detectable systems and leading to higher inferred values of the collapse timescale $t_c$. In contrast, the high-threshold scenario assumes that only very strong signals (e.g., with SNR comparable to GW190425 in a future detector) are included. The detection of even one event above such a high threshold implies a more efficient collapse process, allowing for lower inferred values of $t_c$, assuming a fixed local dark matter density $\rho_\chi$. From these $t_c$ estimates, we then derive the corresponding constraints on $m_\chi$ and $\sigma_\chi$.

An important observation is that the width of the inferred constraints is dominated by the current uncertainty in the local merger rate. At present, this uncertainty is large, which limits the precision of any population-based inference. However, with the advent of more sensitive detector networks, we expect this uncertainty to reduce significantly. Since we cannot reliably forecast the improvement in merger rate measurements, we use current bounds to estimate our constraints for all networks considered. As a result, our forecasts should be viewed as conservative estimates, and can be refined in the future once tighter bounds on the local merger rate become available.

We also note that the detectable merger rates are informed by our best knowledge of the assumed priors on the delay time distributions for BNS ($\propto t_d^{-1} \in [0.02, t_{\rm Hubble}]$ Gyr, along with the star formation rate which inherently have some uncertainty.
Accurate knowledge of the host galaxy, and formation channel with coincident observations will alleviate delay time distribution causing the overall rate to change.
For example,~\citep{Zevin:2022dbo} reports a suppressed BNS merger rate density using delay time distributions informed by the host galaxies of observed short gamma-ray bursts, which would lower the overall detectable rate for merger of $1-3~M_\odot$ objects affecting the $N_{\rm BBH}$ regardless of $t_c$.

Several caveats apply to our analysis. First, our estimates of the collapse timescale $t_c$ rely on assumed values for quantities such as NS temperature and dark matter velocity dispersion. Although these values are astrophysically motivated, variations could shift the resulting dark matter constraints. Second, our results are based on a specific dark matter–induced collapse scenario that produces BHs in the NS mass range. Other mechanisms, such as accretion-induced collapse or primordial BHs~\cite{Carr:2020xqk}, could also produce low-mass black holes. However, these processes are not expected to populate only this specific mass range. Accretion tends to produce more massive remnants, while primordial BHs may span a much broader mass spectrum with their merger rate density peaking at a significantly higher redshift of $\sim 30$. Such differences could help distinguish among these scenarios in future population studies.

Our analysis highlights that GW observations can serve as a complementary tool for constraining dark matter properties, independent of traditional detection methods. The framework developed in this work can be applied to such low-mass merger events in the future, offering a novel use of GW astronomy in probing beyond-standard-model physics.

\section{Acknowledgments}
We thank Bangalore Sathyaprakash for his insightful comments and guidance throughout this study. We thank Pratyusava Baral for the LSC Publication \&
Presentation Committee review of this manuscript. We also thank Koustav Chandra, Ish Gupta, and Viviana Cáceres for their helpful comments. This material is based upon work supported by NSF's LIGO Laboratory which is a major facility fully funded by the National Science Foundation. SK is also supported by NSF awards AST-2307147 and PHY-2308886. SK and DS acknowledge grants OAC-2346596, OAC-2201445, OAC-2103662, OAC-2018299, and PHY-2110594, which provided access to the GWAVE cluster at PennState. DS acknowledges support from NSF Grant PHY-2020275 (Network for Neutrinos, Nuclear Astrophysics, and Symmetries (N3AS)).

\appendix
\section{Event Classification Metrics}
\label{appendix:SD}
The most commonly used metric to quantify support for one model over another is the Bayes factor. In Table~\ref{tab:bayesfactor}, we present $ln \mathcal{BF}^{\mathrm{BBH}}_{\mathrm{BNS}}$, the natural logarithm of the Bayes factor in favor of the BBH hypothesis over the BNS hypothesis.

\begin{table}[b]
\caption{Bayes Factors in favor of the BBH hypothesis over the BNS hypothesis.}
\label{tab:bayesfactor}
\begin{ruledtabular}
\begin{tabular}{lc}
    \textbf{Network} & \textbf{ln$\mathcal{BF}^{BBH}_{BNS}$}  \\
    \hline
    HLVK+ & 5.46\\
    HLI$\#$ & 6.07\\
    CE4020ET &  9.95\\
\end{tabular}
\end{ruledtabular} 
\end{table}
The results indicate that even networks such as HLVK+ are capable of supporting a confident classification between BBH and BNS sources. However, an important caveat is that the BBH waveform model includes two fewer parameters than the BNS model, as it excludes tidal parameters. By Occam’s razor, models with fewer parameters are favored if they fit the data comparably well. Despite lacking explicit tidal terms, the BBH model still captures much of the signal because tidal effects enter at the 5PN order (see Eq.~\ref{eq:5PN}) and contribute primarily at higher frequencies. Consequently, their contribution to the total SNR and thus, impact on the likelihood is relatively minor.

As a result of the reduced parameter space and the subdominant nature of tidal contributions, the Bayes factor $\ln \mathcal{BF}^{\mathrm{BBH}}_{\mathrm{BNS}}$ tends to be biased in favor of the BBH model, even when the true source is a BNS. To address this bias, we consider the Savage–Dickey ratio as an alternative metric for quantifying model support. Table~\ref{tab:results} lists the computed Savage–Dickey ratio for all considered networks in this study.

Let $H_0$ denote the BBH hypothesis, which assumes $\tilde{\Lambda} = 0$, and let $H_1$ denote the BNS hypothesis, under which $\tilde{\Lambda}$ varies freely over the prior range. Under these conditions, the Bayes factor in favor of the BBH model over the BNS model is given by the Savage–Dickey ratio:
$$\mathcal{B}_{\mathrm{BBH}} = \frac{p_1(\tilde{\Lambda} = 0 \mid d)}{\pi_1(\tilde{\Lambda} = 0)},$$
where $p_1(\tilde{\Lambda} = 0 \mid d)$ is the posterior density under the BNS model evaluated at $\tilde{\Lambda} = 0$, and $\pi_1(\tilde{\Lambda} = 0)$ is the corresponding prior density (as also mentioned in \cite{CalderonBustillo:2024akj}). As the inferred posterior for $\tilde{\Lambda}$ has a finite width, we evaluate the Savage-Dickey ratio on a grid of points with $\tilde{\Lambda}<1$. The spread of the resulting ratios is negligible, so we report only the median value in Sec. \ref{sec4A}.

In this context, $p_1$ refers to the family of $\tilde{\Lambda}$ posteriors we compute for all simulated signals, regardless of whether the true source is a BNS or a BBH. As long as the inference is performed using a BNS waveform model, we obtain a posterior on $\tilde{\Lambda}$, and the Savage–Dickey ratio remains well-defined.

For example, in the case of the HLVK+ network, $\mathcal{B}_{\mathrm{BBH}}$ remains modest, indicating that the BBH model is not strongly favored unless the posterior density at $\tilde{\Lambda} = 0$ significantly exceeds the prior. Moreover, we observe that $\mathcal{B}_{\mathrm{BBH}}$ is not highly sensitive to the true nature of the source. Whether the signal arises from a stiff-EoS BNS, a soft-EoS BNS, or an actual BBH, the posterior at $\tilde{\Lambda} = 0$ does not change significantly, reinforcing the need for higher-fidelity data to enable robust classification.

\bibliography{references}{}

\begin{thebibliography}{86}%
\makeatletter
\providecommand \@ifxundefined [1]{%
 \@ifx{#1\undefined}
}%
\providecommand \@ifnum [1]{%
 \ifnum #1\expandafter \@firstoftwo
 \else \expandafter \@secondoftwo
 \fi
}%
\providecommand \@ifx [1]{%
 \ifx #1\expandafter \@firstoftwo
 \else \expandafter \@secondoftwo
 \fi
}%
\providecommand \natexlab [1]{#1}%
\providecommand \enquote  [1]{``#1''}%
\providecommand \bibnamefont  [1]{#1}%
\providecommand \bibfnamefont [1]{#1}%
\providecommand \citenamefont [1]{#1}%
\providecommand \href@noop [0]{\@secondoftwo}%
\providecommand \href [0]{\begingroup \@sanitize@url \@href}%
\providecommand \@href[1]{\@@startlink{#1}\@@href}%
\providecommand \@@href[1]{\endgroup#1\@@endlink}%
\providecommand \@sanitize@url [0]{\catcode `\\12\catcode `\$12\catcode `\&12\catcode `\#12\catcode `\^12\catcode `\_12\catcode `\%12\relax}%
\providecommand \@@startlink[1]{}%
\providecommand \@@endlink[0]{}%
\providecommand \url  [0]{\begingroup\@sanitize@url \@url }%
\providecommand \@url [1]{\endgroup\@href {#1}{\urlprefix }}%
\providecommand \urlprefix  [0]{URL }%
\providecommand \Eprint [0]{\href }%
\providecommand \doibase [0]{https://doi.org/}%
\providecommand \selectlanguage [0]{\@gobble}%
\providecommand \bibinfo  [0]{\@secondoftwo}%
\providecommand \bibfield  [0]{\@secondoftwo}%
\providecommand \translation [1]{[#1]}%
\providecommand \BibitemOpen [0]{}%
\providecommand \bibitemStop [0]{}%
\providecommand \bibitemNoStop [0]{.\EOS\space}%
\providecommand \EOS [0]{\spacefactor3000\relax}%
\providecommand \BibitemShut  [1]{\csname bibitem#1\endcsname}%
\let\auto@bib@innerbib\@empty
\bibitem [{\citenamefont {Abbott}\ \emph {et~al.}(2019{\natexlab{a}})\citenamefont {Abbott} \emph {et~al.}}]{LIGOScientific:2018mvr}%
  \BibitemOpen
  \bibfield  {author} {\bibinfo {author} {\bibfnamefont {B.~P.}\ \bibnamefont {Abbott}} \emph {et~al.} (\bibinfo {collaboration} {LIGO Scientific, Virgo}),\ }\bibfield  {title} {\bibinfo {title} {{GWTC-1: A Gravitational-Wave Transient Catalog of Compact Binary Mergers Observed by LIGO and Virgo during the First and Second Observing Runs}},\ }\href {https://doi.org/10.1103/PhysRevX.9.031040} {\bibfield  {journal} {\bibinfo  {journal} {Phys. Rev. X}\ }\textbf {\bibinfo {volume} {9}},\ \bibinfo {pages} {031040} (\bibinfo {year} {2019}{\natexlab{a}})},\ \Eprint {https://arxiv.org/abs/1811.12907} {arXiv:1811.12907 [astro-ph.HE]} \BibitemShut {NoStop}%
\bibitem [{\citenamefont {Abbott}\ \emph {et~al.}(2021{\natexlab{a}})\citenamefont {Abbott} \emph {et~al.}}]{LIGOScientific:2020ibl}%
  \BibitemOpen
  \bibfield  {author} {\bibinfo {author} {\bibfnamefont {R.}~\bibnamefont {Abbott}} \emph {et~al.} (\bibinfo {collaboration} {LIGO Scientific, Virgo}),\ }\bibfield  {title} {\bibinfo {title} {{GWTC-2: Compact Binary Coalescences Observed by LIGO and Virgo During the First Half of the Third Observing Run}},\ }\href {https://doi.org/10.1103/PhysRevX.11.021053} {\bibfield  {journal} {\bibinfo  {journal} {Phys. Rev. X}\ }\textbf {\bibinfo {volume} {11}},\ \bibinfo {pages} {021053} (\bibinfo {year} {2021}{\natexlab{a}})},\ \Eprint {https://arxiv.org/abs/2010.14527} {arXiv:2010.14527 [gr-qc]} \BibitemShut {NoStop}%
\bibitem [{\citenamefont {Abbott}\ \emph {et~al.}(2023{\natexlab{a}})\citenamefont {Abbott} \emph {et~al.}}]{KAGRA:2021vkt}%
  \BibitemOpen
  \bibfield  {author} {\bibinfo {author} {\bibfnamefont {R.}~\bibnamefont {Abbott}} \emph {et~al.} (\bibinfo {collaboration} {KAGRA, VIRGO, LIGO Scientific}),\ }\bibfield  {title} {\bibinfo {title} {{GWTC-3: Compact Binary Coalescences Observed by LIGO and Virgo during the Second Part of the Third Observing Run}},\ }\href {https://doi.org/10.1103/PhysRevX.13.041039} {\bibfield  {journal} {\bibinfo  {journal} {Phys. Rev. X}\ }\textbf {\bibinfo {volume} {13}},\ \bibinfo {pages} {041039} (\bibinfo {year} {2023}{\natexlab{a}})},\ \Eprint {https://arxiv.org/abs/2111.03606} {arXiv:2111.03606 [gr-qc]} \BibitemShut {NoStop}%
\bibitem [{\citenamefont {Abbott}\ \emph {et~al.}(2017{\natexlab{a}})\citenamefont {Abbott} \emph {et~al.}}]{LIGOScientific:2017vwq}%
  \BibitemOpen
  \bibfield  {author} {\bibinfo {author} {\bibfnamefont {B.~P.}\ \bibnamefont {Abbott}} \emph {et~al.} (\bibinfo {collaboration} {LIGO Scientific, Virgo}),\ }\bibfield  {title} {\bibinfo {title} {{GW170817: Observation of Gravitational Waves from a Binary Neutron Star Inspiral}},\ }\href {https://doi.org/10.1103/PhysRevLett.119.161101} {\bibfield  {journal} {\bibinfo  {journal} {Phys. Rev. Lett.}\ }\textbf {\bibinfo {volume} {119}},\ \bibinfo {pages} {161101} (\bibinfo {year} {2017}{\natexlab{a}})},\ \Eprint {https://arxiv.org/abs/1710.05832} {arXiv:1710.05832 [gr-qc]} \BibitemShut {NoStop}%
\bibitem [{\citenamefont {Abbott}\ \emph {et~al.}(2019{\natexlab{b}})\citenamefont {Abbott} \emph {et~al.}}]{LIGOScientific:2018hze}%
  \BibitemOpen
  \bibfield  {author} {\bibinfo {author} {\bibfnamefont {B.~P.}\ \bibnamefont {Abbott}} \emph {et~al.} (\bibinfo {collaboration} {LIGO Scientific, Virgo}),\ }\bibfield  {title} {\bibinfo {title} {{Properties of the binary neutron star merger GW170817}},\ }\href {https://doi.org/10.1103/PhysRevX.9.011001} {\bibfield  {journal} {\bibinfo  {journal} {Phys. Rev. X}\ }\textbf {\bibinfo {volume} {9}},\ \bibinfo {pages} {011001} (\bibinfo {year} {2019}{\natexlab{b}})},\ \Eprint {https://arxiv.org/abs/1805.11579} {arXiv:1805.11579 [gr-qc]} \BibitemShut {NoStop}%
\bibitem [{\citenamefont {Abbott}\ \emph {et~al.}(2018{\natexlab{a}})\citenamefont {Abbott} \emph {et~al.}}]{LIGOScientific:2018cki}%
  \BibitemOpen
  \bibfield  {author} {\bibinfo {author} {\bibfnamefont {B.~P.}\ \bibnamefont {Abbott}} \emph {et~al.} (\bibinfo {collaboration} {LIGO Scientific, Virgo}),\ }\bibfield  {title} {\bibinfo {title} {{GW170817: Measurements of neutron star radii and equation of state}},\ }\href {https://doi.org/10.1103/PhysRevLett.121.161101} {\bibfield  {journal} {\bibinfo  {journal} {Phys. Rev. Lett.}\ }\textbf {\bibinfo {volume} {121}},\ \bibinfo {pages} {161101} (\bibinfo {year} {2018}{\natexlab{a}})},\ \Eprint {https://arxiv.org/abs/1805.11581} {arXiv:1805.11581 [gr-qc]} \BibitemShut {NoStop}%
\bibitem [{\citenamefont {Coulter}\ \emph {et~al.}(2017)\citenamefont {Coulter} \emph {et~al.}}]{Coulter:2017wya}%
  \BibitemOpen
  \bibfield  {author} {\bibinfo {author} {\bibfnamefont {D.~A.}\ \bibnamefont {Coulter}} \emph {et~al.},\ }\bibfield  {title} {\bibinfo {title} {{Swope Supernova Survey 2017a (SSS17a), the Optical Counterpart to a Gravitational Wave Source}},\ }\href {https://doi.org/10.1126/science.aap9811} {\bibfield  {journal} {\bibinfo  {journal} {Science}\ }\textbf {\bibinfo {volume} {358}},\ \bibinfo {pages} {1556} (\bibinfo {year} {2017})},\ \Eprint {https://arxiv.org/abs/1710.05452} {arXiv:1710.05452 [astro-ph.HE]} \BibitemShut {NoStop}%
\bibitem [{\citenamefont {Soares-Santos}\ \emph {et~al.}(2017)\citenamefont {Soares-Santos} \emph {et~al.}}]{DES:2017kbs}%
  \BibitemOpen
  \bibfield  {author} {\bibinfo {author} {\bibfnamefont {M.}~\bibnamefont {Soares-Santos}} \emph {et~al.} (\bibinfo {collaboration} {DES, Dark Energy Camera GW-EM}),\ }\bibfield  {title} {\bibinfo {title} {{The Electromagnetic Counterpart of the Binary Neutron Star Merger LIGO/Virgo GW170817. I. Discovery of the Optical Counterpart Using the Dark Energy Camera}},\ }\href {https://doi.org/10.3847/2041-8213/aa9059} {\bibfield  {journal} {\bibinfo  {journal} {Astrophys. J. Lett.}\ }\textbf {\bibinfo {volume} {848}},\ \bibinfo {pages} {L16} (\bibinfo {year} {2017})},\ \Eprint {https://arxiv.org/abs/1710.05459} {arXiv:1710.05459 [astro-ph.HE]} \BibitemShut {NoStop}%
\bibitem [{\citenamefont {Abbott}\ \emph {et~al.}(2017{\natexlab{b}})\citenamefont {Abbott} \emph {et~al.}}]{LIGOScientific:2017zic}%
  \BibitemOpen
  \bibfield  {author} {\bibinfo {author} {\bibfnamefont {B.~P.}\ \bibnamefont {Abbott}} \emph {et~al.} (\bibinfo {collaboration} {LIGO Scientific, Virgo, Fermi-GBM, INTEGRAL}),\ }\bibfield  {title} {\bibinfo {title} {{Gravitational Waves and Gamma-rays from a Binary Neutron Star Merger: GW170817 and GRB 170817A}},\ }\href {https://doi.org/10.3847/2041-8213/aa920c} {\bibfield  {journal} {\bibinfo  {journal} {Astrophys. J. Lett.}\ }\textbf {\bibinfo {volume} {848}},\ \bibinfo {pages} {L13} (\bibinfo {year} {2017}{\natexlab{b}})},\ \Eprint {https://arxiv.org/abs/1710.05834} {arXiv:1710.05834 [astro-ph.HE]} \BibitemShut {NoStop}%
\bibitem [{\citenamefont {Goldstein}\ \emph {et~al.}(2017)\citenamefont {Goldstein} \emph {et~al.}}]{Goldstein:2017mmi}%
  \BibitemOpen
  \bibfield  {author} {\bibinfo {author} {\bibfnamefont {A.}~\bibnamefont {Goldstein}} \emph {et~al.},\ }\bibfield  {title} {\bibinfo {title} {{An Ordinary Short Gamma-Ray Burst with Extraordinary Implications: Fermi-GBM Detection of GRB 170817A}},\ }\href {https://doi.org/10.3847/2041-8213/aa8f41} {\bibfield  {journal} {\bibinfo  {journal} {Astrophys. J. Lett.}\ }\textbf {\bibinfo {volume} {848}},\ \bibinfo {pages} {L14} (\bibinfo {year} {2017})},\ \Eprint {https://arxiv.org/abs/1710.05446} {arXiv:1710.05446 [astro-ph.HE]} \BibitemShut {NoStop}%
\bibitem [{\citenamefont {Savchenko}\ \emph {et~al.}(2017)\citenamefont {Savchenko} \emph {et~al.}}]{Savchenko:2017ffs}%
  \BibitemOpen
  \bibfield  {author} {\bibinfo {author} {\bibfnamefont {V.}~\bibnamefont {Savchenko}} \emph {et~al.},\ }\bibfield  {title} {\bibinfo {title} {{INTEGRAL Detection of the First Prompt Gamma-Ray Signal Coincident with the Gravitational-wave Event GW170817}},\ }\href {https://doi.org/10.3847/2041-8213/aa8f94} {\bibfield  {journal} {\bibinfo  {journal} {Astrophys. J. Lett.}\ }\textbf {\bibinfo {volume} {848}},\ \bibinfo {pages} {L15} (\bibinfo {year} {2017})},\ \Eprint {https://arxiv.org/abs/1710.05449} {arXiv:1710.05449 [astro-ph.HE]} \BibitemShut {NoStop}%
\bibitem [{\citenamefont {Hinderer}\ \emph {et~al.}(2019)\citenamefont {Hinderer} \emph {et~al.}}]{Hinderer:2018pei}%
  \BibitemOpen
  \bibfield  {author} {\bibinfo {author} {\bibfnamefont {T.}~\bibnamefont {Hinderer}} \emph {et~al.},\ }\bibfield  {title} {\bibinfo {title} {{Distinguishing the nature of comparable-mass neutron star binary systems with multimessenger observations: GW170817 case study}},\ }\href {https://doi.org/10.1103/PhysRevD.100.063021} {\bibfield  {journal} {\bibinfo  {journal} {Phys. Rev. D}\ }\textbf {\bibinfo {volume} {100}},\ \bibinfo {pages} {06321} (\bibinfo {year} {2019})},\ \Eprint {https://arxiv.org/abs/1808.03836} {arXiv:1808.03836 [astro-ph.HE]} \BibitemShut {NoStop}%
\bibitem [{\citenamefont {Coughlin}\ and\ \citenamefont {Dietrich}(2019)}]{Coughlin:2019kqf}%
  \BibitemOpen
  \bibfield  {author} {\bibinfo {author} {\bibfnamefont {M.~W.}\ \bibnamefont {Coughlin}}\ and\ \bibinfo {author} {\bibfnamefont {T.}~\bibnamefont {Dietrich}},\ }\bibfield  {title} {\bibinfo {title} {{Can a black hole\textendash{}neutron star merger explain GW170817, AT2017gfo, and GRB170817A?}},\ }\href {https://doi.org/10.1103/PhysRevD.100.043011} {\bibfield  {journal} {\bibinfo  {journal} {Phys. Rev. D}\ }\textbf {\bibinfo {volume} {100}},\ \bibinfo {pages} {043011} (\bibinfo {year} {2019})},\ \Eprint {https://arxiv.org/abs/1901.06052} {arXiv:1901.06052 [astro-ph.HE]} \BibitemShut {NoStop}%
\bibitem [{\citenamefont {Abbott}\ \emph {et~al.}(2020{\natexlab{a}})\citenamefont {Abbott} \emph {et~al.}}]{LIGOScientific:2020aai}%
  \BibitemOpen
  \bibfield  {author} {\bibinfo {author} {\bibfnamefont {B.~P.}\ \bibnamefont {Abbott}} \emph {et~al.} (\bibinfo {collaboration} {LIGO Scientific, Virgo}),\ }\bibfield  {title} {\bibinfo {title} {{GW190425: Observation of a Compact Binary Coalescence with Total Mass $\sim 3.4 M_{\odot}$}},\ }\href {https://doi.org/10.3847/2041-8213/ab75f5} {\bibfield  {journal} {\bibinfo  {journal} {Astrophys. J. Lett.}\ }\textbf {\bibinfo {volume} {892}},\ \bibinfo {pages} {L3} (\bibinfo {year} {2020}{\natexlab{a}})},\ \Eprint {https://arxiv.org/abs/2001.01761} {arXiv:2001.01761 [astro-ph.HE]} \BibitemShut {NoStop}%
\bibitem [{\citenamefont {Damour}\ and\ \citenamefont {Nagar}(2009)}]{Damour:2009vw}%
  \BibitemOpen
  \bibfield  {author} {\bibinfo {author} {\bibfnamefont {T.}~\bibnamefont {Damour}}\ and\ \bibinfo {author} {\bibfnamefont {A.}~\bibnamefont {Nagar}},\ }\bibfield  {title} {\bibinfo {title} {{Relativistic tidal properties of neutron stars}},\ }\href {https://doi.org/10.1103/PhysRevD.80.084035} {\bibfield  {journal} {\bibinfo  {journal} {Phys. Rev. D}\ }\textbf {\bibinfo {volume} {80}},\ \bibinfo {pages} {084035} (\bibinfo {year} {2009})},\ \Eprint {https://arxiv.org/abs/0906.0096} {arXiv:0906.0096 [gr-qc]} \BibitemShut {NoStop}%
\bibitem [{\citenamefont {Binnington}\ and\ \citenamefont {Poisson}(2009)}]{Binnington:2009bb}%
  \BibitemOpen
  \bibfield  {author} {\bibinfo {author} {\bibfnamefont {T.}~\bibnamefont {Binnington}}\ and\ \bibinfo {author} {\bibfnamefont {E.}~\bibnamefont {Poisson}},\ }\bibfield  {title} {\bibinfo {title} {{Relativistic theory of tidal Love numbers}},\ }\href {https://doi.org/10.1103/PhysRevD.80.084018} {\bibfield  {journal} {\bibinfo  {journal} {Phys. Rev. D}\ }\textbf {\bibinfo {volume} {80}},\ \bibinfo {pages} {084018} (\bibinfo {year} {2009})},\ \Eprint {https://arxiv.org/abs/0906.1366} {arXiv:0906.1366 [gr-qc]} \BibitemShut {NoStop}%
\bibitem [{\citenamefont {Rhoades}\ and\ \citenamefont {Ruffini}(1974)}]{Rhoades:1974fn}%
  \BibitemOpen
  \bibfield  {author} {\bibinfo {author} {\bibfnamefont {C.~E.}\ \bibnamefont {Rhoades}, \bibfnamefont {Jr.}}\ and\ \bibinfo {author} {\bibfnamefont {R.}~\bibnamefont {Ruffini}},\ }\bibfield  {title} {\bibinfo {title} {{Maximum mass of a neutron star}},\ }\href {https://doi.org/10.1103/PhysRevLett.32.324} {\bibfield  {journal} {\bibinfo  {journal} {Phys. Rev. Lett.}\ }\textbf {\bibinfo {volume} {32}},\ \bibinfo {pages} {324} (\bibinfo {year} {1974})}\BibitemShut {NoStop}%
\bibitem [{\citenamefont {Kalogera}\ and\ \citenamefont {Baym}(1996)}]{Kalogera:1996ci}%
  \BibitemOpen
  \bibfield  {author} {\bibinfo {author} {\bibfnamefont {V.}~\bibnamefont {Kalogera}}\ and\ \bibinfo {author} {\bibfnamefont {G.}~\bibnamefont {Baym}},\ }\bibfield  {title} {\bibinfo {title} {{The maximum mass of a neutron star}},\ }\href {https://doi.org/10.1086/310296} {\bibfield  {journal} {\bibinfo  {journal} {Astrophys. J. Lett.}\ }\textbf {\bibinfo {volume} {470}},\ \bibinfo {pages} {L61} (\bibinfo {year} {1996})},\ \Eprint {https://arxiv.org/abs/astro-ph/9608059} {arXiv:astro-ph/9608059} \BibitemShut {NoStop}%
\bibitem [{\citenamefont {Bailyn}\ \emph {et~al.}(1998)\citenamefont {Bailyn}, \citenamefont {Jain}, \citenamefont {Coppi},\ and\ \citenamefont {Orosz}}]{Bailyn:1997xt}%
  \BibitemOpen
  \bibfield  {author} {\bibinfo {author} {\bibfnamefont {C.~D.}\ \bibnamefont {Bailyn}}, \bibinfo {author} {\bibfnamefont {R.~K.}\ \bibnamefont {Jain}}, \bibinfo {author} {\bibfnamefont {P.}~\bibnamefont {Coppi}},\ and\ \bibinfo {author} {\bibfnamefont {J.~A.}\ \bibnamefont {Orosz}},\ }\bibfield  {title} {\bibinfo {title} {{The Mass distribution of stellar black holes}},\ }\href {https://doi.org/10.1086/305614} {\bibfield  {journal} {\bibinfo  {journal} {Astrophys. J.}\ }\textbf {\bibinfo {volume} {499}},\ \bibinfo {pages} {367} (\bibinfo {year} {1998})},\ \Eprint {https://arxiv.org/abs/astro-ph/9708032} {arXiv:astro-ph/9708032} \BibitemShut {NoStop}%
\bibitem [{\citenamefont {Ozel}\ \emph {et~al.}(2010)\citenamefont {Ozel}, \citenamefont {Psaltis}, \citenamefont {Narayan},\ and\ \citenamefont {McClintock}}]{Ozel:2010su}%
  \BibitemOpen
  \bibfield  {author} {\bibinfo {author} {\bibfnamefont {F.}~\bibnamefont {Ozel}}, \bibinfo {author} {\bibfnamefont {D.}~\bibnamefont {Psaltis}}, \bibinfo {author} {\bibfnamefont {R.}~\bibnamefont {Narayan}},\ and\ \bibinfo {author} {\bibfnamefont {J.~E.}\ \bibnamefont {McClintock}},\ }\bibfield  {title} {\bibinfo {title} {{The Black Hole Mass Distribution in the Galaxy}},\ }\href {https://doi.org/10.1088/0004-637X/725/2/1918} {\bibfield  {journal} {\bibinfo  {journal} {Astrophys. J.}\ }\textbf {\bibinfo {volume} {725}},\ \bibinfo {pages} {1918} (\bibinfo {year} {2010})},\ \Eprint {https://arxiv.org/abs/1006.2834} {arXiv:1006.2834 [astro-ph.GA]} \BibitemShut {NoStop}%
\bibitem [{\citenamefont {Farr}\ \emph {et~al.}(2011)\citenamefont {Farr}, \citenamefont {Sravan}, \citenamefont {Cantrell}, \citenamefont {Kreidberg}, \citenamefont {Bailyn}, \citenamefont {Mandel},\ and\ \citenamefont {Kalogera}}]{Farr:2010tu}%
  \BibitemOpen
  \bibfield  {author} {\bibinfo {author} {\bibfnamefont {W.~M.}\ \bibnamefont {Farr}}, \bibinfo {author} {\bibfnamefont {N.}~\bibnamefont {Sravan}}, \bibinfo {author} {\bibfnamefont {A.}~\bibnamefont {Cantrell}}, \bibinfo {author} {\bibfnamefont {L.}~\bibnamefont {Kreidberg}}, \bibinfo {author} {\bibfnamefont {C.~D.}\ \bibnamefont {Bailyn}}, \bibinfo {author} {\bibfnamefont {I.}~\bibnamefont {Mandel}},\ and\ \bibinfo {author} {\bibfnamefont {V.}~\bibnamefont {Kalogera}},\ }\bibfield  {title} {\bibinfo {title} {{The Mass Distribution of Stellar-Mass Black Holes}},\ }\href {https://doi.org/10.1088/0004-637X/741/2/103} {\bibfield  {journal} {\bibinfo  {journal} {Astrophys. J.}\ }\textbf {\bibinfo {volume} {741}},\ \bibinfo {pages} {103} (\bibinfo {year} {2011})},\ \Eprint {https://arxiv.org/abs/1011.1459} {arXiv:1011.1459 [astro-ph.GA]} \BibitemShut {NoStop}%
\bibitem [{\citenamefont {Thompson}\ \emph {et~al.}(2018)\citenamefont {Thompson} \emph {et~al.}}]{Thompson:2018ycv}%
  \BibitemOpen
  \bibfield  {author} {\bibinfo {author} {\bibfnamefont {T.~A.}\ \bibnamefont {Thompson}} \emph {et~al.},\ }\bibfield  {title} {\bibinfo {title} {{Discovery of a Candidate Black Hole - Giant Star Binary System in the Galactic Field}},\ }\bibfield  {journal} {\bibinfo  {journal} {Science}\ }\href {https://doi.org/10.1126/science.aau4005} {10.1126/science.aau4005} (\bibinfo {year} {2018}),\ \Eprint {https://arxiv.org/abs/1806.02751} {arXiv:1806.02751 [astro-ph.HE]} \BibitemShut {NoStop}%
\bibitem [{\citenamefont {Jayasinghe}\ \emph {et~al.}(2021)\citenamefont {Jayasinghe} \emph {et~al.}}]{Jayasinghe:2021uqb}%
  \BibitemOpen
  \bibfield  {author} {\bibinfo {author} {\bibfnamefont {T.}~\bibnamefont {Jayasinghe}} \emph {et~al.},\ }\bibfield  {title} {\bibinfo {title} {{A unicorn in monoceros: the 3\,M\ensuremath{\odot} dark companion to the bright, nearby red giant V723 Mon is a non-interacting, mass-gap black hole candidate}},\ }\href {https://doi.org/10.1093/mnras/stab907} {\bibfield  {journal} {\bibinfo  {journal} {Mon. Not. Roy. Astron. Soc.}\ }\textbf {\bibinfo {volume} {504}},\ \bibinfo {pages} {2577} (\bibinfo {year} {2021})},\ \Eprint {https://arxiv.org/abs/2101.02212} {arXiv:2101.02212 [astro-ph.SR]} \BibitemShut {NoStop}%
\bibitem [{\citenamefont {Lam}\ \emph {et~al.}(2022)\citenamefont {Lam} \emph {et~al.}}]{Lam:2022vuq}%
  \BibitemOpen
  \bibfield  {author} {\bibinfo {author} {\bibfnamefont {C.~Y.}\ \bibnamefont {Lam}} \emph {et~al.},\ }\bibfield  {title} {\bibinfo {title} {{An Isolated Mass-gap Black Hole or Neutron Star Detected with Astrometric Microlensing}},\ }\href {https://doi.org/10.3847/2041-8213/ac7442} {\bibfield  {journal} {\bibinfo  {journal} {Astrophys. J. Lett.}\ }\textbf {\bibinfo {volume} {933}},\ \bibinfo {pages} {L23} (\bibinfo {year} {2022})},\ \Eprint {https://arxiv.org/abs/2202.01903} {arXiv:2202.01903 [astro-ph.GA]} \BibitemShut {NoStop}%
\bibitem [{\citenamefont {Abac}\ \emph {et~al.}(2024{\natexlab{a}})\citenamefont {Abac} \emph {et~al.}}]{LIGOScientific:2024elc}%
  \BibitemOpen
  \bibfield  {author} {\bibinfo {author} {\bibfnamefont {A.~G.}\ \bibnamefont {Abac}} \emph {et~al.} (\bibinfo {collaboration} {LIGO Scientific, Virgo,, KAGRA, VIRGO}),\ }\bibfield  {title} {\bibinfo {title} {{Observation of Gravitational Waves from the Coalescence of a 2.5\textendash{}4.5 M $_{\odot}$ Compact Object and a Neutron Star}},\ }\href {https://doi.org/10.3847/2041-8213/ad5beb} {\bibfield  {journal} {\bibinfo  {journal} {Astrophys. J. Lett.}\ }\textbf {\bibinfo {volume} {970}},\ \bibinfo {pages} {L34} (\bibinfo {year} {2024}{\natexlab{a}})},\ \Eprint {https://arxiv.org/abs/2404.04248} {arXiv:2404.04248 [astro-ph.HE]} \BibitemShut {NoStop}%
\bibitem [{\citenamefont {Abbott}\ \emph {et~al.}(2021{\natexlab{b}})\citenamefont {Abbott} \emph {et~al.}}]{LIGOScientific:2021qlt}%
  \BibitemOpen
  \bibfield  {author} {\bibinfo {author} {\bibfnamefont {R.}~\bibnamefont {Abbott}} \emph {et~al.} (\bibinfo {collaboration} {LIGO Scientific, KAGRA, VIRGO}),\ }\bibfield  {title} {\bibinfo {title} {{Observation of Gravitational Waves from Two Neutron Star\textendash{}Black Hole Coalescences}},\ }\href {https://doi.org/10.3847/2041-8213/ac082e} {\bibfield  {journal} {\bibinfo  {journal} {Astrophys. J. Lett.}\ }\textbf {\bibinfo {volume} {915}},\ \bibinfo {pages} {L5} (\bibinfo {year} {2021}{\natexlab{b}})},\ \Eprint {https://arxiv.org/abs/2106.15163} {arXiv:2106.15163 [astro-ph.HE]} \BibitemShut {NoStop}%
\bibitem [{\citenamefont {Abbott}\ \emph {et~al.}(2020{\natexlab{b}})\citenamefont {Abbott} \emph {et~al.}}]{LIGOScientific:2020zkf}%
  \BibitemOpen
  \bibfield  {author} {\bibinfo {author} {\bibfnamefont {R.}~\bibnamefont {Abbott}} \emph {et~al.} (\bibinfo {collaboration} {LIGO Scientific, Virgo}),\ }\bibfield  {title} {\bibinfo {title} {{GW190814: Gravitational Waves from the Coalescence of a 23 Solar Mass Black Hole with a 2.6 Solar Mass Compact Object}},\ }\href {https://doi.org/10.3847/2041-8213/ab960f} {\bibfield  {journal} {\bibinfo  {journal} {Astrophys. J. Lett.}\ }\textbf {\bibinfo {volume} {896}},\ \bibinfo {pages} {L44} (\bibinfo {year} {2020}{\natexlab{b}})},\ \Eprint {https://arxiv.org/abs/2006.12611} {arXiv:2006.12611 [astro-ph.HE]} \BibitemShut {NoStop}%
\bibitem [{\citenamefont {Singh}\ \emph {et~al.}(2023)\citenamefont {Singh}, \citenamefont {Gupta}, \citenamefont {Berti}, \citenamefont {Reddy},\ and\ \citenamefont {Sathyaprakash}}]{Singh:2022wvw}%
  \BibitemOpen
  \bibfield  {author} {\bibinfo {author} {\bibfnamefont {D.}~\bibnamefont {Singh}}, \bibinfo {author} {\bibfnamefont {A.}~\bibnamefont {Gupta}}, \bibinfo {author} {\bibfnamefont {E.}~\bibnamefont {Berti}}, \bibinfo {author} {\bibfnamefont {S.}~\bibnamefont {Reddy}},\ and\ \bibinfo {author} {\bibfnamefont {B.~S.}\ \bibnamefont {Sathyaprakash}},\ }\bibfield  {title} {\bibinfo {title} {{Constraining properties of asymmetric dark matter candidates from gravitational-wave observations}},\ }\href {https://doi.org/10.1103/PhysRevD.107.083037} {\bibfield  {journal} {\bibinfo  {journal} {Phys. Rev. D}\ }\textbf {\bibinfo {volume} {107}},\ \bibinfo {pages} {083037} (\bibinfo {year} {2023})},\ \Eprint {https://arxiv.org/abs/2210.15739} {arXiv:2210.15739 [gr-qc]} \BibitemShut {NoStop}%
\bibitem [{\citenamefont {Barsotti}\ \emph {et~al.}(2018)\citenamefont {Barsotti}, \citenamefont {McCuller}, \citenamefont {Evans},\ and\ \citenamefont {Fritschel}}]{T1800042}%
  \BibitemOpen
  \bibfield  {author} {\bibinfo {author} {\bibfnamefont {L.}~\bibnamefont {Barsotti}}, \bibinfo {author} {\bibfnamefont {L.}~\bibnamefont {McCuller}}, \bibinfo {author} {\bibfnamefont {M.}~\bibnamefont {Evans}},\ and\ \bibinfo {author} {\bibfnamefont {P.}~\bibnamefont {Fritschel}},\ }\href {https://dcc.ligo.org/public/0149/T1800042/004/T1800042-v4.pdf} {\emph {\bibinfo {title} {The A+ Design Curve}}},\ \bibinfo {type} {Tech. Rep.}\ \bibinfo {number} {T1800042}\ (\bibinfo  {institution} {LIGO},\ \bibinfo {year} {2018})\BibitemShut {NoStop}%
\bibitem [{\citenamefont {Fritschel}\ \emph {et~al.}(2022)\citenamefont {Fritschel}, \citenamefont {Kuns}, \citenamefont {Driggers}, \citenamefont {Effler}, \citenamefont {Lantz}, \citenamefont {Ottaway}, \citenamefont {Ballmer}, \citenamefont {Dooley}, \citenamefont {Adhikari}, \citenamefont {Evans}, \citenamefont {Farr}, \citenamefont {Gonzalez}, \citenamefont {Schmidt},\ and\ \citenamefont {Raja}}]{T2200287}%
  \BibitemOpen
  \bibfield  {author} {\bibinfo {author} {\bibfnamefont {P.}~\bibnamefont {Fritschel}}, \bibinfo {author} {\bibfnamefont {K.}~\bibnamefont {Kuns}}, \bibinfo {author} {\bibfnamefont {J.}~\bibnamefont {Driggers}}, \bibinfo {author} {\bibfnamefont {A.}~\bibnamefont {Effler}}, \bibinfo {author} {\bibfnamefont {B.}~\bibnamefont {Lantz}}, \bibinfo {author} {\bibfnamefont {D.}~\bibnamefont {Ottaway}}, \bibinfo {author} {\bibfnamefont {S.}~\bibnamefont {Ballmer}}, \bibinfo {author} {\bibfnamefont {K.}~\bibnamefont {Dooley}}, \bibinfo {author} {\bibfnamefont {R.}~\bibnamefont {Adhikari}}, \bibinfo {author} {\bibfnamefont {M.}~\bibnamefont {Evans}}, \bibinfo {author} {\bibfnamefont {B.}~\bibnamefont {Farr}}, \bibinfo {author} {\bibfnamefont {G.}~\bibnamefont {Gonzalez}}, \bibinfo {author} {\bibfnamefont {P.}~\bibnamefont {Schmidt}},\ and\ \bibinfo {author} {\bibfnamefont {S.}~\bibnamefont {Raja}},\ }\href {https://dcc.ligo.org/LIGO-T2200287/public} {\emph {\bibinfo {title} {Report from the LSC Post-O5 Study Group}}},\
  \bibinfo {type} {Tech. Rep.}\ \bibinfo {number} {T2200287}\ (\bibinfo  {institution} {LIGO},\ \bibinfo {year} {2022})\BibitemShut {NoStop}%
\bibitem [{\citenamefont {Reitze}\ \emph {et~al.}(2019)\citenamefont {Reitze} \emph {et~al.}}]{Reitze:2019iox}%
  \BibitemOpen
  \bibfield  {author} {\bibinfo {author} {\bibfnamefont {D.}~\bibnamefont {Reitze}} \emph {et~al.},\ }\bibfield  {title} {\bibinfo {title} {{Cosmic Explorer: The U.S. Contribution to Gravitational-Wave Astronomy beyond LIGO}},\ }\href@noop {} {\bibfield  {journal} {\bibinfo  {journal} {Bull. Am. Astron. Soc.}\ }\textbf {\bibinfo {volume} {51}},\ \bibinfo {pages} {035} (\bibinfo {year} {2019})},\ \Eprint {https://arxiv.org/abs/1907.04833} {arXiv:1907.04833 [astro-ph.IM]} \BibitemShut {NoStop}%
\bibitem [{\citenamefont {Evans}\ \emph {et~al.}(2021)\citenamefont {Evans} \emph {et~al.}}]{Evans:2021gyd}%
  \BibitemOpen
  \bibfield  {author} {\bibinfo {author} {\bibfnamefont {M.}~\bibnamefont {Evans}} \emph {et~al.},\ }\href@noop {} {\bibinfo {title} {{A Horizon Study for Cosmic Explorer: Science, Observatories, and Community}}} (\bibinfo {year} {2021}),\ \Eprint {https://arxiv.org/abs/2109.09882} {arXiv:2109.09882 [astro-ph.IM]} \BibitemShut {NoStop}%
\bibitem [{\citenamefont {Abbott}\ \emph {et~al.}(2017{\natexlab{c}})\citenamefont {Abbott} \emph {et~al.}}]{LIGOScientific:2016wof}%
  \BibitemOpen
  \bibfield  {author} {\bibinfo {author} {\bibfnamefont {B.~P.}\ \bibnamefont {Abbott}} \emph {et~al.} (\bibinfo {collaboration} {LIGO Scientific}),\ }\bibfield  {title} {\bibinfo {title} {{Exploring the Sensitivity of Next Generation Gravitational Wave Detectors}},\ }\href {https://doi.org/10.1088/1361-6382/aa51f4} {\bibfield  {journal} {\bibinfo  {journal} {Class. Quant. Grav.}\ }\textbf {\bibinfo {volume} {34}},\ \bibinfo {pages} {044001} (\bibinfo {year} {2017}{\natexlab{c}})},\ \Eprint {https://arxiv.org/abs/1607.08697} {arXiv:1607.08697 [astro-ph.IM]} \BibitemShut {NoStop}%
\bibitem [{\citenamefont {Punturo}\ \emph {et~al.}(2010)\citenamefont {Punturo} \emph {et~al.}}]{Punturo:2010zza}%
  \BibitemOpen
  \bibfield  {author} {\bibinfo {author} {\bibfnamefont {M.}~\bibnamefont {Punturo}} \emph {et~al.},\ }\bibfield  {title} {\bibinfo {title} {{The third generation of gravitational wave observatories and their science reach}},\ }\href {https://doi.org/10.1088/0264-9381/27/8/084007} {\bibfield  {journal} {\bibinfo  {journal} {Class. Quant. Grav.}\ }\textbf {\bibinfo {volume} {27}},\ \bibinfo {pages} {084007} (\bibinfo {year} {2010})}\BibitemShut {NoStop}%
\bibitem [{\citenamefont {Hild}\ \emph {et~al.}(2011)\citenamefont {Hild} \emph {et~al.}}]{Hild:2010id}%
  \BibitemOpen
  \bibfield  {author} {\bibinfo {author} {\bibfnamefont {S.}~\bibnamefont {Hild}} \emph {et~al.},\ }\bibfield  {title} {\bibinfo {title} {{Sensitivity Studies for Third-Generation Gravitational Wave Observatories}},\ }\href {https://doi.org/10.1088/0264-9381/28/9/094013} {\bibfield  {journal} {\bibinfo  {journal} {Class. Quant. Grav.}\ }\textbf {\bibinfo {volume} {28}},\ \bibinfo {pages} {094013} (\bibinfo {year} {2011})},\ \Eprint {https://arxiv.org/abs/1012.0908} {arXiv:1012.0908 [gr-qc]} \BibitemShut {NoStop}%
\bibitem [{\citenamefont {Maggiore}\ \emph {et~al.}(2020)\citenamefont {Maggiore} \emph {et~al.}}]{ET:2019dnz}%
  \BibitemOpen
  \bibfield  {author} {\bibinfo {author} {\bibfnamefont {M.}~\bibnamefont {Maggiore}} \emph {et~al.} (\bibinfo {collaboration} {ET}),\ }\bibfield  {title} {\bibinfo {title} {{Science Case for the Einstein Telescope}},\ }\href {https://doi.org/10.1088/1475-7516/2020/03/050} {\bibfield  {journal} {\bibinfo  {journal} {JCAP}\ }\textbf {\bibinfo {volume} {03}}\bibfield  {number} {\bibinfo  {number} { (3)},\ \bibinfo {pages} {050}},\ }\Eprint {https://arxiv.org/abs/1912.02622} {arXiv:1912.02622 [astro-ph.CO]} \BibitemShut {NoStop}%
\bibitem [{\citenamefont {Collaboration}\ and\ \citenamefont {Collaboration}(2022{\natexlab{a}})}]{LIGOScientific:2022}%
  \BibitemOpen
  \bibfield  {author} {\bibinfo {author} {\bibfnamefont {L.~S.}\ \bibnamefont {Collaboration}}\ and\ \bibinfo {author} {\bibfnamefont {V.}~\bibnamefont {Collaboration}},\ }\href {https://doi.org/10.5281/zenodo.6513631} {\bibinfo {title} {Gwtc-2.1: Deep extended catalog of compact binary coalescences observed by ligo and virgo during the first half of the third observing run - parameter estimation data release (version v2)}} (\bibinfo {year} {2022}{\natexlab{a}})\BibitemShut {NoStop}%
\bibitem [{\citenamefont {Abbott}\ \emph {et~al.}(2021{\natexlab{c}})\citenamefont {Abbott} \emph {et~al.}}]{LIGOScientific:2019lzm}%
  \BibitemOpen
  \bibfield  {author} {\bibinfo {author} {\bibfnamefont {R.}~\bibnamefont {Abbott}} \emph {et~al.} (\bibinfo {collaboration} {LIGO Scientific, Virgo}),\ }\bibfield  {title} {\bibinfo {title} {{Open data from the first and second observing runs of Advanced LIGO and Advanced Virgo}},\ }\href {https://doi.org/10.1016/j.softx.2021.100658} {\bibfield  {journal} {\bibinfo  {journal} {SoftwareX}\ }\textbf {\bibinfo {volume} {13}},\ \bibinfo {pages} {100658} (\bibinfo {year} {2021}{\natexlab{c}})},\ \Eprint {https://arxiv.org/abs/1912.11716} {arXiv:1912.11716 [gr-qc]} \BibitemShut {NoStop}%
\bibitem [{\citenamefont {Abbott}\ \emph {et~al.}(2023{\natexlab{b}})\citenamefont {Abbott} \emph {et~al.}}]{KAGRA:2023pio}%
  \BibitemOpen
  \bibfield  {author} {\bibinfo {author} {\bibfnamefont {R.}~\bibnamefont {Abbott}} \emph {et~al.} (\bibinfo {collaboration} {KAGRA, VIRGO, LIGO Scientific}),\ }\bibfield  {title} {\bibinfo {title} {{Open Data from the Third Observing Run of LIGO, Virgo, KAGRA, and GEO}},\ }\href {https://doi.org/10.3847/1538-4365/acdc9f} {\bibfield  {journal} {\bibinfo  {journal} {Astrophys. J. Suppl.}\ }\textbf {\bibinfo {volume} {267}},\ \bibinfo {pages} {29} (\bibinfo {year} {2023}{\natexlab{b}})},\ \Eprint {https://arxiv.org/abs/2302.03676} {arXiv:2302.03676 [gr-qc]} \BibitemShut {NoStop}%
\bibitem [{\citenamefont {Pratten}\ \emph {et~al.}(2020)\citenamefont {Pratten}, \citenamefont {Husa}, \citenamefont {Garcia-Quiros}, \citenamefont {Colleoni}, \citenamefont {Ramos-Buades}, \citenamefont {Estelles},\ and\ \citenamefont {Jaume}}]{Pratten:2020fqn}%
  \BibitemOpen
  \bibfield  {author} {\bibinfo {author} {\bibfnamefont {G.}~\bibnamefont {Pratten}}, \bibinfo {author} {\bibfnamefont {S.}~\bibnamefont {Husa}}, \bibinfo {author} {\bibfnamefont {C.}~\bibnamefont {Garcia-Quiros}}, \bibinfo {author} {\bibfnamefont {M.}~\bibnamefont {Colleoni}}, \bibinfo {author} {\bibfnamefont {A.}~\bibnamefont {Ramos-Buades}}, \bibinfo {author} {\bibfnamefont {H.}~\bibnamefont {Estelles}},\ and\ \bibinfo {author} {\bibfnamefont {R.}~\bibnamefont {Jaume}},\ }\bibfield  {title} {\bibinfo {title} {{Setting the cornerstone for a family of models for gravitational waves from compact binaries: The dominant harmonic for nonprecessing quasicircular black holes}},\ }\href {https://doi.org/10.1103/PhysRevD.102.064001} {\bibfield  {journal} {\bibinfo  {journal} {Phys. Rev. D}\ }\textbf {\bibinfo {volume} {102}},\ \bibinfo {pages} {064001} (\bibinfo {year} {2020})},\ \Eprint {https://arxiv.org/abs/2001.11412} {arXiv:2001.11412 [gr-qc]} \BibitemShut {NoStop}%
\bibitem [{\citenamefont {Abac}\ \emph {et~al.}(2024{\natexlab{b}})\citenamefont {Abac}, \citenamefont {Dietrich}, \citenamefont {Buonanno}, \citenamefont {Steinhoff},\ and\ \citenamefont {Ujevic}}]{Abac:2023ujg}%
  \BibitemOpen
  \bibfield  {author} {\bibinfo {author} {\bibfnamefont {A.}~\bibnamefont {Abac}}, \bibinfo {author} {\bibfnamefont {T.}~\bibnamefont {Dietrich}}, \bibinfo {author} {\bibfnamefont {A.}~\bibnamefont {Buonanno}}, \bibinfo {author} {\bibfnamefont {J.}~\bibnamefont {Steinhoff}},\ and\ \bibinfo {author} {\bibfnamefont {M.}~\bibnamefont {Ujevic}},\ }\bibfield  {title} {\bibinfo {title} {{New and robust gravitational-waveform model for high-mass-ratio binary neutron star systems with dynamical tidal effects}},\ }\href {https://doi.org/10.1103/PhysRevD.109.024062} {\bibfield  {journal} {\bibinfo  {journal} {Phys. Rev. D}\ }\textbf {\bibinfo {volume} {109}},\ \bibinfo {pages} {024062} (\bibinfo {year} {2024}{\natexlab{b}})},\ \Eprint {https://arxiv.org/abs/2311.07456} {arXiv:2311.07456 [gr-qc]} \BibitemShut {NoStop}%
\bibitem [{\citenamefont {Dickey}(1971)}]{Dickey1971}%
  \BibitemOpen
  \bibfield  {author} {\bibinfo {author} {\bibfnamefont {J.~M.}\ \bibnamefont {Dickey}},\ }\bibfield  {title} {\bibinfo {title} {The weighted likelihood ratio, linear hypotheses on normal location parameters},\ }\href {https://doi.org/10.1214/aoms/1177693507} {\bibfield  {journal} {\bibinfo  {journal} {The Annals of Mathematical Statistics}\ }\textbf {\bibinfo {volume} {42}},\ \bibinfo {pages} {204} (\bibinfo {year} {1971})}\BibitemShut {NoStop}%
\bibitem [{\citenamefont {Hinderer}(2008)}]{Hinderer:2007mb}%
  \BibitemOpen
  \bibfield  {author} {\bibinfo {author} {\bibfnamefont {T.}~\bibnamefont {Hinderer}},\ }\bibfield  {title} {\bibinfo {title} {{Tidal Love numbers of neutron stars}},\ }\href {https://doi.org/10.1086/533487} {\bibfield  {journal} {\bibinfo  {journal} {Astrophys. J.}\ }\textbf {\bibinfo {volume} {677}},\ \bibinfo {pages} {1216} (\bibinfo {year} {2008})},\ \bibinfo {note} {[Erratum: Astrophys.J. 697, 964 (2009)]},\ \Eprint {https://arxiv.org/abs/0711.2420} {arXiv:0711.2420 [astro-ph]} \BibitemShut {NoStop}%
\bibitem [{\citenamefont {Flanagan}\ and\ \citenamefont {Hinderer}(2008)}]{Flanagan:2007ix}%
  \BibitemOpen
  \bibfield  {author} {\bibinfo {author} {\bibfnamefont {E.~E.}\ \bibnamefont {Flanagan}}\ and\ \bibinfo {author} {\bibfnamefont {T.}~\bibnamefont {Hinderer}},\ }\bibfield  {title} {\bibinfo {title} {{Constraining neutron star tidal Love numbers with gravitational wave detectors}},\ }\href {https://doi.org/10.1103/PhysRevD.77.021502} {\bibfield  {journal} {\bibinfo  {journal} {Phys. Rev. D}\ }\textbf {\bibinfo {volume} {77}},\ \bibinfo {pages} {021502} (\bibinfo {year} {2008})},\ \Eprint {https://arxiv.org/abs/0709.1915} {arXiv:0709.1915 [astro-ph]} \BibitemShut {NoStop}%
\bibitem [{\citenamefont {Sathyaprakash}\ and\ \citenamefont {Dhurandhar}(1991)}]{PhysRevD.44.3819}%
  \BibitemOpen
  \bibfield  {author} {\bibinfo {author} {\bibfnamefont {B.~S.}\ \bibnamefont {Sathyaprakash}}\ and\ \bibinfo {author} {\bibfnamefont {S.~V.}\ \bibnamefont {Dhurandhar}},\ }\bibfield  {title} {\bibinfo {title} {Choice of filters for the detection of gravitational waves from coalescing binaries},\ }\href {https://doi.org/10.1103/PhysRevD.44.3819} {\bibfield  {journal} {\bibinfo  {journal} {Phys. Rev. D}\ }\textbf {\bibinfo {volume} {44}},\ \bibinfo {pages} {3819} (\bibinfo {year} {1991})}\BibitemShut {NoStop}%
\bibitem [{\citenamefont {Favata}(2014)}]{Favata:2013rwa}%
  \BibitemOpen
  \bibfield  {author} {\bibinfo {author} {\bibfnamefont {M.}~\bibnamefont {Favata}},\ }\bibfield  {title} {\bibinfo {title} {{Systematic parameter errors in inspiraling neutron star binaries}},\ }\href {https://doi.org/10.1103/PhysRevLett.112.101101} {\bibfield  {journal} {\bibinfo  {journal} {Phys. Rev. Lett.}\ }\textbf {\bibinfo {volume} {112}},\ \bibinfo {pages} {101101} (\bibinfo {year} {2014})},\ \Eprint {https://arxiv.org/abs/1310.8288} {arXiv:1310.8288 [gr-qc]} \BibitemShut {NoStop}%
\bibitem [{\citenamefont {Abbott}\ \emph {et~al.}(2024)\citenamefont {Abbott} \emph {et~al.}}]{LIGOScientific:2021usb}%
  \BibitemOpen
  \bibfield  {author} {\bibinfo {author} {\bibfnamefont {R.}~\bibnamefont {Abbott}} \emph {et~al.} (\bibinfo {collaboration} {LIGO Scientific, VIRGO}),\ }\bibfield  {title} {\bibinfo {title} {{GWTC-2.1: Deep extended catalog of compact binary coalescences observed by LIGO and Virgo during the first half of the third observing run}},\ }\href {https://doi.org/10.1103/PhysRevD.109.022001} {\bibfield  {journal} {\bibinfo  {journal} {Phys. Rev. D}\ }\textbf {\bibinfo {volume} {109}},\ \bibinfo {pages} {022001} (\bibinfo {year} {2024})},\ \Eprint {https://arxiv.org/abs/2108.01045} {arXiv:2108.01045 [gr-qc]} \BibitemShut {NoStop}%
\bibitem [{\citenamefont {Romero-Shaw}\ \emph {et~al.}(2020)\citenamefont {Romero-Shaw} \emph {et~al.}}]{Romero-Shaw:2020owr}%
  \BibitemOpen
  \bibfield  {author} {\bibinfo {author} {\bibfnamefont {I.~M.}\ \bibnamefont {Romero-Shaw}} \emph {et~al.},\ }\bibfield  {title} {\bibinfo {title} {{Bayesian inference for compact binary coalescences with bilby: validation and application to the first LIGO\textendash{}Virgo gravitational-wave transient catalogue}},\ }\href {https://doi.org/10.1093/mnras/staa2850} {\bibfield  {journal} {\bibinfo  {journal} {Mon. Not. Roy. Astron. Soc.}\ }\textbf {\bibinfo {volume} {499}},\ \bibinfo {pages} {3295} (\bibinfo {year} {2020})},\ \Eprint {https://arxiv.org/abs/2006.00714} {arXiv:2006.00714 [astro-ph.IM]} \BibitemShut {NoStop}%
\bibitem [{\citenamefont {Ashton}\ \emph {et~al.}(2019)\citenamefont {Ashton} \emph {et~al.}}]{Ashton:2018jfp}%
  \BibitemOpen
  \bibfield  {author} {\bibinfo {author} {\bibfnamefont {G.}~\bibnamefont {Ashton}} \emph {et~al.},\ }\bibfield  {title} {\bibinfo {title} {{BILBY: A user-friendly Bayesian inference library for gravitational-wave astronomy}},\ }\href {https://doi.org/10.3847/1538-4365/ab06fc} {\bibfield  {journal} {\bibinfo  {journal} {Astrophys. J. Suppl.}\ }\textbf {\bibinfo {volume} {241}},\ \bibinfo {pages} {27} (\bibinfo {year} {2019})},\ \Eprint {https://arxiv.org/abs/1811.02042} {arXiv:1811.02042 [astro-ph.IM]} \BibitemShut {NoStop}%
\bibitem [{\citenamefont {{LIGO Scientific Collaboration}}\ \emph {et~al.}(2018)\citenamefont {{LIGO Scientific Collaboration}}, \citenamefont {{Virgo Collaboration}},\ and\ \citenamefont {{KAGRA Collaboration}}}]{lalsuite}%
  \BibitemOpen
  \bibfield  {author} {\bibinfo {author} {\bibnamefont {{LIGO Scientific Collaboration}}}, \bibinfo {author} {\bibnamefont {{Virgo Collaboration}}},\ and\ \bibinfo {author} {\bibnamefont {{KAGRA Collaboration}}},\ }\href {https://doi.org/10.7935/GT1W-FZ16} {\bibinfo {title} {{LVK} {A}lgorithm {L}ibrary - {LALS}uite}},\ \bibinfo {howpublished} {Free software (GPL)} (\bibinfo {year} {2018})\BibitemShut {NoStop}%
\bibitem [{\citenamefont {Wette}(2020)}]{swiglal}%
  \BibitemOpen
  \bibfield  {author} {\bibinfo {author} {\bibfnamefont {K.}~\bibnamefont {Wette}},\ }\bibfield  {title} {\bibinfo {title} {{SWIGLAL: Python and Octave interfaces to the LALSuite gravitational-wave data analysis libraries}},\ }\href {https://doi.org/10.1016/j.softx.2020.100634} {\bibfield  {journal} {\bibinfo  {journal} {SoftwareX}\ }\textbf {\bibinfo {volume} {12}},\ \bibinfo {pages} {100634} (\bibinfo {year} {2020})}\BibitemShut {NoStop}%
\bibitem [{\citenamefont {Speagle}(2020)}]{Speagle:2019ivv}%
  \BibitemOpen
  \bibfield  {author} {\bibinfo {author} {\bibfnamefont {J.~S.}\ \bibnamefont {Speagle}},\ }\bibfield  {title} {\bibinfo {title} {{dynesty: a dynamic nested sampling package for estimating Bayesian posteriors and evidences}},\ }\href {https://doi.org/10.1093/mnras/staa278} {\bibfield  {journal} {\bibinfo  {journal} {Mon. Not. Roy. Astron. Soc.}\ }\textbf {\bibinfo {volume} {493}},\ \bibinfo {pages} {3132} (\bibinfo {year} {2020})},\ \Eprint {https://arxiv.org/abs/1904.02180} {arXiv:1904.02180 [astro-ph.IM]} \BibitemShut {NoStop}%
\bibitem [{\citenamefont {Cornish}(2010)}]{Cornish:2010kf}%
  \BibitemOpen
  \bibfield  {author} {\bibinfo {author} {\bibfnamefont {N.~J.}\ \bibnamefont {Cornish}},\ }\href@noop {} {\bibinfo {title} {{Fast Fisher Matrices and Lazy Likelihoods}}} (\bibinfo {year} {2010}),\ \Eprint {https://arxiv.org/abs/1007.4820} {arXiv:1007.4820 [gr-qc]} \BibitemShut {NoStop}%
\bibitem [{\citenamefont {Cornish}(2021)}]{Cornish:2021lje}%
  \BibitemOpen
  \bibfield  {author} {\bibinfo {author} {\bibfnamefont {N.~J.}\ \bibnamefont {Cornish}},\ }\bibfield  {title} {\bibinfo {title} {{Heterodyned likelihood for rapid gravitational wave parameter inference}},\ }\href {https://doi.org/10.1103/PhysRevD.104.104054} {\bibfield  {journal} {\bibinfo  {journal} {Phys. Rev. D}\ }\textbf {\bibinfo {volume} {104}},\ \bibinfo {pages} {104054} (\bibinfo {year} {2021})},\ \Eprint {https://arxiv.org/abs/2109.02728} {arXiv:2109.02728 [gr-qc]} \BibitemShut {NoStop}%
\bibitem [{\citenamefont {Zackay}\ \emph {et~al.}(2018)\citenamefont {Zackay}, \citenamefont {Dai},\ and\ \citenamefont {Venumadhav}}]{Zackay:2018qdy}%
  \BibitemOpen
  \bibfield  {author} {\bibinfo {author} {\bibfnamefont {B.}~\bibnamefont {Zackay}}, \bibinfo {author} {\bibfnamefont {L.}~\bibnamefont {Dai}},\ and\ \bibinfo {author} {\bibfnamefont {T.}~\bibnamefont {Venumadhav}},\ }\href@noop {} {\bibinfo {title} {{Relative Binning and Fast Likelihood Evaluation for Gravitational Wave Parameter Estimation}}} (\bibinfo {year} {2018}),\ \Eprint {https://arxiv.org/abs/1806.08792} {arXiv:1806.08792 [astro-ph.IM]} \BibitemShut {NoStop}%
\bibitem [{\citenamefont {Antoniadis}\ \emph {et~al.}(2016)\citenamefont {Antoniadis}, \citenamefont {Tauris}, \citenamefont {Ozel}, \citenamefont {Barr}, \citenamefont {Champion},\ and\ \citenamefont {Freire}}]{Antoniadis:2016hxz}%
  \BibitemOpen
  \bibfield  {author} {\bibinfo {author} {\bibfnamefont {J.}~\bibnamefont {Antoniadis}}, \bibinfo {author} {\bibfnamefont {T.~M.}\ \bibnamefont {Tauris}}, \bibinfo {author} {\bibfnamefont {F.}~\bibnamefont {Ozel}}, \bibinfo {author} {\bibfnamefont {E.}~\bibnamefont {Barr}}, \bibinfo {author} {\bibfnamefont {D.~J.}\ \bibnamefont {Champion}},\ and\ \bibinfo {author} {\bibfnamefont {P.~C.~C.}\ \bibnamefont {Freire}},\ }\href@noop {} {\bibinfo {title} {{The millisecond pulsar mass distribution: Evidence for bimodality and constraints on the maximum neutron star mass}}} (\bibinfo {year} {2016}),\ \Eprint {https://arxiv.org/abs/1605.01665} {arXiv:1605.01665 [astro-ph.HE]} \BibitemShut {NoStop}%
\bibitem [{\citenamefont {Alsing}\ \emph {et~al.}(2018)\citenamefont {Alsing}, \citenamefont {Silva},\ and\ \citenamefont {Berti}}]{Alsing:2017bbc}%
  \BibitemOpen
  \bibfield  {author} {\bibinfo {author} {\bibfnamefont {J.}~\bibnamefont {Alsing}}, \bibinfo {author} {\bibfnamefont {H.~O.}\ \bibnamefont {Silva}},\ and\ \bibinfo {author} {\bibfnamefont {E.}~\bibnamefont {Berti}},\ }\bibfield  {title} {\bibinfo {title} {{Evidence for a maximum mass cut-off in the neutron star mass distribution and constraints on the equation of state}},\ }\href {https://doi.org/10.1093/mnras/sty1065} {\bibfield  {journal} {\bibinfo  {journal} {Mon. Not. Roy. Astron. Soc.}\ }\textbf {\bibinfo {volume} {478}},\ \bibinfo {pages} {1377} (\bibinfo {year} {2018})},\ \Eprint {https://arxiv.org/abs/1709.07889} {arXiv:1709.07889 [astro-ph.HE]} \BibitemShut {NoStop}%
\bibitem [{\citenamefont {{Farr}}\ and\ \citenamefont {{Chatziioannou}}(2020)}]{2020RNAAS...4...65F}%
  \BibitemOpen
  \bibfield  {author} {\bibinfo {author} {\bibfnamefont {W.~M.}\ \bibnamefont {{Farr}}}\ and\ \bibinfo {author} {\bibfnamefont {K.}~\bibnamefont {{Chatziioannou}}},\ }\bibfield  {title} {\bibinfo {title} {{A Population-Informed Mass Estimate for Pulsar J0740+6620}},\ }\href {https://doi.org/10.3847/2515-5172/ab9088} {\bibfield  {journal} {\bibinfo  {journal} {Research Notes of the American Astronomical Society}\ }\textbf {\bibinfo {volume} {4}},\ \bibinfo {eid} {65} (\bibinfo {year} {2020})},\ \Eprint {https://arxiv.org/abs/2005.00032} {arXiv:2005.00032 [astro-ph.GA]} \BibitemShut {NoStop}%
\bibitem [{\citenamefont {Shao}\ \emph {et~al.}(2020)\citenamefont {Shao}, \citenamefont {Tang}, \citenamefont {Jiang},\ and\ \citenamefont {Fan}}]{Shao:2020bzt}%
  \BibitemOpen
  \bibfield  {author} {\bibinfo {author} {\bibfnamefont {D.-S.}\ \bibnamefont {Shao}}, \bibinfo {author} {\bibfnamefont {S.-P.}\ \bibnamefont {Tang}}, \bibinfo {author} {\bibfnamefont {J.-L.}\ \bibnamefont {Jiang}},\ and\ \bibinfo {author} {\bibfnamefont {Y.-Z.}\ \bibnamefont {Fan}},\ }\bibfield  {title} {\bibinfo {title} {{Maximum mass cutoff in the neutron star mass distribution and the prospect of forming supramassive objects in the double neutron star mergers}},\ }\href {https://doi.org/10.1103/PhysRevD.102.063006} {\bibfield  {journal} {\bibinfo  {journal} {Phys. Rev. D}\ }\textbf {\bibinfo {volume} {102}},\ \bibinfo {pages} {063006} (\bibinfo {year} {2020})},\ \Eprint {https://arxiv.org/abs/2009.04275} {arXiv:2009.04275 [astro-ph.HE]} \BibitemShut {NoStop}%
\bibitem [{\citenamefont {Landry}\ and\ \citenamefont {Read}(2021)}]{Landry:2021hvl}%
  \BibitemOpen
  \bibfield  {author} {\bibinfo {author} {\bibfnamefont {P.}~\bibnamefont {Landry}}\ and\ \bibinfo {author} {\bibfnamefont {J.~S.}\ \bibnamefont {Read}},\ }\bibfield  {title} {\bibinfo {title} {{The Mass Distribution of Neutron Stars in Gravitational-wave Binaries}},\ }\href {https://doi.org/10.3847/2041-8213/ac2f3e} {\bibfield  {journal} {\bibinfo  {journal} {Astrophys. J. Lett.}\ }\textbf {\bibinfo {volume} {921}},\ \bibinfo {pages} {L25} (\bibinfo {year} {2021})},\ \Eprint {https://arxiv.org/abs/2107.04559} {arXiv:2107.04559 [astro-ph.HE]} \BibitemShut {NoStop}%
\bibitem [{\citenamefont {{European Gravitational Observatory}}(2022)}]{ET-0304B-22}%
  \BibitemOpen
  \bibfield  {author} {\bibinfo {author} {\bibnamefont {{European Gravitational Observatory}}},\ }\href {https://apps.et-gw.eu/tds/?call_file=ET-0304B-22_ETSensitivityCurvesUsedForCoBA.pdf} {\bibinfo {title} {Et-0304b-22: Etsensitivitycurvesusedforcoba}} (\bibinfo {year} {2022}),\ \bibinfo {note} {accessed: 2024-12-19}\BibitemShut {NoStop}%
\bibitem [{\citenamefont {Abbott}\ \emph {et~al.}(2018{\natexlab{b}})\citenamefont {Abbott} \emph {et~al.}}]{KAGRA:2013rdx}%
  \BibitemOpen
  \bibfield  {author} {\bibinfo {author} {\bibfnamefont {B.~P.}\ \bibnamefont {Abbott}} \emph {et~al.} (\bibinfo {collaboration} {KAGRA, LIGO Scientific, Virgo, VIRGO}),\ }\bibfield  {title} {\bibinfo {title} {{Prospects for observing and localizing gravitational-wave transients with Advanced LIGO, Advanced Virgo and KAGRA}},\ }\href {https://doi.org/10.1007/s41114-020-00026-9} {\bibfield  {journal} {\bibinfo  {journal} {Living Rev. Rel.}\ }\textbf {\bibinfo {volume} {21}},\ \bibinfo {pages} {3} (\bibinfo {year} {2018}{\natexlab{b}})},\ \Eprint {https://arxiv.org/abs/1304.0670} {arXiv:1304.0670 [gr-qc]} \BibitemShut {NoStop}%
\bibitem [{\citenamefont {Aasi}\ \emph {et~al.}(2015)\citenamefont {Aasi} \emph {et~al.}}]{LIGOScientific:2014pky}%
  \BibitemOpen
  \bibfield  {author} {\bibinfo {author} {\bibfnamefont {J.}~\bibnamefont {Aasi}} \emph {et~al.} (\bibinfo {collaboration} {LIGO Scientific}),\ }\bibfield  {title} {\bibinfo {title} {{Advanced LIGO}},\ }\href {https://doi.org/10.1088/0264-9381/32/7/074001} {\bibfield  {journal} {\bibinfo  {journal} {Class. Quant. Grav.}\ }\textbf {\bibinfo {volume} {32}},\ \bibinfo {pages} {074001} (\bibinfo {year} {2015})},\ \Eprint {https://arxiv.org/abs/1411.4547} {arXiv:1411.4547 [gr-qc]} \BibitemShut {NoStop}%
\bibitem [{\citenamefont {Akutsu}\ \emph {et~al.}(2020)\citenamefont {Akutsu} \emph {et~al.}}]{10.1093/ptep/ptaa125}%
  \BibitemOpen
  \bibfield  {author} {\bibinfo {author} {\bibfnamefont {T.}~\bibnamefont {Akutsu}} \emph {et~al.},\ }\bibfield  {title} {\bibinfo {title} {Overview of kagra: Detector design and construction history},\ }\href {https://doi.org/10.1093/ptep/ptaa125} {\bibfield  {journal} {\bibinfo  {journal} {Progress of Theoretical and Experimental Physics}\ }\textbf {\bibinfo {volume} {2021}},\ \bibinfo {pages} {05A101} (\bibinfo {year} {2020})},\ \Eprint {https://arxiv.org/abs/https://academic.oup.com/ptep/article-pdf/2021/5/05A101/37974994/ptaa125.pdf} {https://academic.oup.com/ptep/article-pdf/2021/5/05A101/37974994/ptaa125.pdf} \BibitemShut {NoStop}%
\bibitem [{\citenamefont {Acernese}\ \emph {et~al.}(2015)\citenamefont {Acernese} \emph {et~al.}}]{VIRGO:2014yos}%
  \BibitemOpen
  \bibfield  {author} {\bibinfo {author} {\bibfnamefont {F.}~\bibnamefont {Acernese}} \emph {et~al.} (\bibinfo {collaboration} {VIRGO}),\ }\bibfield  {title} {\bibinfo {title} {{Advanced Virgo: a second-generation interferometric gravitational wave detector}},\ }\href {https://doi.org/10.1088/0264-9381/32/2/024001} {\bibfield  {journal} {\bibinfo  {journal} {Class. Quant. Grav.}\ }\textbf {\bibinfo {volume} {32}},\ \bibinfo {pages} {024001} (\bibinfo {year} {2015})},\ \Eprint {https://arxiv.org/abs/1408.3978} {arXiv:1408.3978 [gr-qc]} \BibitemShut {NoStop}%
\bibitem [{\citenamefont {Saleem}\ \emph {et~al.}(2022)\citenamefont {Saleem} \emph {et~al.}}]{Saleem:2021iwi}%
  \BibitemOpen
  \bibfield  {author} {\bibinfo {author} {\bibfnamefont {M.}~\bibnamefont {Saleem}} \emph {et~al.},\ }\bibfield  {title} {\bibinfo {title} {{The science case for LIGO-India}},\ }\href {https://doi.org/10.1088/1361-6382/ac3b99} {\bibfield  {journal} {\bibinfo  {journal} {Class. Quant. Grav.}\ }\textbf {\bibinfo {volume} {39}},\ \bibinfo {pages} {025004} (\bibinfo {year} {2022})},\ \Eprint {https://arxiv.org/abs/2105.01716} {arXiv:2105.01716 [gr-qc]} \BibitemShut {NoStop}%
\bibitem [{\citenamefont {Collaboration}\ and\ \citenamefont {Collaboration}(2022{\natexlab{b}})}]{ligo_scientific_collaboration_and_virgo_2022_6513631}%
  \BibitemOpen
  \bibfield  {author} {\bibinfo {author} {\bibfnamefont {L.~S.}\ \bibnamefont {Collaboration}}\ and\ \bibinfo {author} {\bibfnamefont {V.}~\bibnamefont {Collaboration}},\ }\bibfield  {title} {\bibinfo {title} {Gwtc-2.1: Deep extended catalog of compact binary coalescences observed by ligo and virgo during the first half of the third observing run - parameter estimation data release},\ }\href {https://doi.org/10.5281/zenodo.6513631} {10.5281/zenodo.6513631} (\bibinfo {year} {2022}{\natexlab{b}})\BibitemShut {NoStop}%
\bibitem [{\citenamefont {Akmal}\ \emph {et~al.}(1998)\citenamefont {Akmal}, \citenamefont {Pandharipande},\ and\ \citenamefont {Ravenhall}}]{Akmal:1998cf}%
  \BibitemOpen
  \bibfield  {author} {\bibinfo {author} {\bibfnamefont {A.}~\bibnamefont {Akmal}}, \bibinfo {author} {\bibfnamefont {V.~R.}\ \bibnamefont {Pandharipande}},\ and\ \bibinfo {author} {\bibfnamefont {D.~G.}\ \bibnamefont {Ravenhall}},\ }\bibfield  {title} {\bibinfo {title} {{The Equation of state of nucleon matter and neutron star structure}},\ }\href {https://doi.org/10.1103/PhysRevC.58.1804} {\bibfield  {journal} {\bibinfo  {journal} {Phys. Rev. C}\ }\textbf {\bibinfo {volume} {58}},\ \bibinfo {pages} {1804} (\bibinfo {year} {1998})},\ \Eprint {https://arxiv.org/abs/nucl-th/9804027} {arXiv:nucl-th/9804027} \BibitemShut {NoStop}%
\bibitem [{\citenamefont {M{\"u}ther}\ \emph {et~al.}(1987)\citenamefont {M{\"u}ther}, \citenamefont {Prakash},\ and\ \citenamefont {Ainsworth}}]{Muther:1987xaa}%
  \BibitemOpen
  \bibfield  {author} {\bibinfo {author} {\bibfnamefont {H.}~\bibnamefont {M{\"u}ther}}, \bibinfo {author} {\bibfnamefont {M.}~\bibnamefont {Prakash}},\ and\ \bibinfo {author} {\bibfnamefont {T.~L.}\ \bibnamefont {Ainsworth}},\ }\bibfield  {title} {\bibinfo {title} {{The nuclear symmetry energy in relativistic Brueckner-Hartree-Fock calculations}},\ }\href {https://doi.org/10.1016/0370-2693(87)91611-X} {\bibfield  {journal} {\bibinfo  {journal} {Phys. Lett. B}\ }\textbf {\bibinfo {volume} {199}},\ \bibinfo {pages} {469} (\bibinfo {year} {1987})}\BibitemShut {NoStop}%
\bibitem [{\citenamefont {Typel}\ \emph {et~al.}(2010)\citenamefont {Typel}, \citenamefont {Ropke}, \citenamefont {Klahn}, \citenamefont {Blaschke},\ and\ \citenamefont {Wolter}}]{Typel:2009sy}%
  \BibitemOpen
  \bibfield  {author} {\bibinfo {author} {\bibfnamefont {S.}~\bibnamefont {Typel}}, \bibinfo {author} {\bibfnamefont {G.}~\bibnamefont {Ropke}}, \bibinfo {author} {\bibfnamefont {T.}~\bibnamefont {Klahn}}, \bibinfo {author} {\bibfnamefont {D.}~\bibnamefont {Blaschke}},\ and\ \bibinfo {author} {\bibfnamefont {H.~H.}\ \bibnamefont {Wolter}},\ }\bibfield  {title} {\bibinfo {title} {{Composition and thermodynamics of nuclear matter with light clusters}},\ }\href {https://doi.org/10.1103/PhysRevC.81.015803} {\bibfield  {journal} {\bibinfo  {journal} {Phys. Rev. C}\ }\textbf {\bibinfo {volume} {81}},\ \bibinfo {pages} {015803} (\bibinfo {year} {2010})},\ \Eprint {https://arxiv.org/abs/0908.2344} {arXiv:0908.2344 [nucl-th]} \BibitemShut {NoStop}%
\bibitem [{\citenamefont {Hempel}\ and\ \citenamefont {Schaffner-Bielich}(2010)}]{Hempel:2009mc}%
  \BibitemOpen
  \bibfield  {author} {\bibinfo {author} {\bibfnamefont {M.}~\bibnamefont {Hempel}}\ and\ \bibinfo {author} {\bibfnamefont {J.}~\bibnamefont {Schaffner-Bielich}},\ }\bibfield  {title} {\bibinfo {title} {{Statistical Model for a Complete Supernova Equation of State}},\ }\href {https://doi.org/10.1016/j.nuclphysa.2010.02.010} {\bibfield  {journal} {\bibinfo  {journal} {Nucl. Phys. A}\ }\textbf {\bibinfo {volume} {837}},\ \bibinfo {pages} {210} (\bibinfo {year} {2010})},\ \Eprint {https://arxiv.org/abs/0911.4073} {arXiv:0911.4073 [nucl-th]} \BibitemShut {NoStop}%
\bibitem [{\citenamefont {Cutler}\ and\ \citenamefont {Flanagan}(1994)}]{Cutler:1994ys}%
  \BibitemOpen
  \bibfield  {author} {\bibinfo {author} {\bibfnamefont {C.}~\bibnamefont {Cutler}}\ and\ \bibinfo {author} {\bibfnamefont {E.~E.}\ \bibnamefont {Flanagan}},\ }\bibfield  {title} {\bibinfo {title} {{Gravitational waves from merging compact binaries: How accurately can one extract the binary's parameters from the inspiral wave form?}},\ }\href {https://doi.org/10.1103/PhysRevD.49.2658} {\bibfield  {journal} {\bibinfo  {journal} {Phys. Rev. D}\ }\textbf {\bibinfo {volume} {49}},\ \bibinfo {pages} {2658} (\bibinfo {year} {1994})},\ \Eprint {https://arxiv.org/abs/gr-qc/9402014} {arXiv:gr-qc/9402014} \BibitemShut {NoStop}%
\bibitem [{\citenamefont {Poisson}\ and\ \citenamefont {Will}(1995)}]{Poisson:1995ef}%
  \BibitemOpen
  \bibfield  {author} {\bibinfo {author} {\bibfnamefont {E.}~\bibnamefont {Poisson}}\ and\ \bibinfo {author} {\bibfnamefont {C.~M.}\ \bibnamefont {Will}},\ }\bibfield  {title} {\bibinfo {title} {{Gravitational waves from inspiraling compact binaries: Parameter estimation using second postNewtonian wave forms}},\ }\href {https://doi.org/10.1103/PhysRevD.52.848} {\bibfield  {journal} {\bibinfo  {journal} {Phys. Rev. D}\ }\textbf {\bibinfo {volume} {52}},\ \bibinfo {pages} {848} (\bibinfo {year} {1995})},\ \Eprint {https://arxiv.org/abs/gr-qc/9502040} {arXiv:gr-qc/9502040} \BibitemShut {NoStop}%
\bibitem [{\citenamefont {Abbott}\ \emph {et~al.}(2023{\natexlab{c}})\citenamefont {Abbott} \emph {et~al.}}]{KAGRA:2021duu}%
  \BibitemOpen
  \bibfield  {author} {\bibinfo {author} {\bibfnamefont {R.}~\bibnamefont {Abbott}} \emph {et~al.} (\bibinfo {collaboration} {KAGRA, VIRGO, LIGO Scientific}),\ }\bibfield  {title} {\bibinfo {title} {{Population of Merging Compact Binaries Inferred Using Gravitational Waves through GWTC-3}},\ }\href {https://doi.org/10.1103/PhysRevX.13.011048} {\bibfield  {journal} {\bibinfo  {journal} {Phys. Rev. X}\ }\textbf {\bibinfo {volume} {13}},\ \bibinfo {pages} {011048} (\bibinfo {year} {2023}{\natexlab{c}})},\ \Eprint {https://arxiv.org/abs/2111.03634} {arXiv:2111.03634 [astro-ph.HE]} \BibitemShut {NoStop}%
\bibitem [{\citenamefont {Madau}\ and\ \citenamefont {Dickinson}(2014)}]{Madau:2014bja}%
  \BibitemOpen
  \bibfield  {author} {\bibinfo {author} {\bibfnamefont {P.}~\bibnamefont {Madau}}\ and\ \bibinfo {author} {\bibfnamefont {M.}~\bibnamefont {Dickinson}},\ }\bibfield  {title} {\bibinfo {title} {{Cosmic Star Formation History}},\ }\href {https://doi.org/10.1146/annurev-astro-081811-125615} {\bibfield  {journal} {\bibinfo  {journal} {Ann. Rev. Astron. Astrophys.}\ }\textbf {\bibinfo {volume} {52}},\ \bibinfo {pages} {415} (\bibinfo {year} {2014})},\ \Eprint {https://arxiv.org/abs/1403.0007} {arXiv:1403.0007 [astro-ph.CO]} \BibitemShut {NoStop}%
\bibitem [{\citenamefont {Gupta}\ \emph {et~al.}(2024)\citenamefont {Gupta} \emph {et~al.}}]{Gupta:2023lga}%
  \BibitemOpen
  \bibfield  {author} {\bibinfo {author} {\bibfnamefont {I.}~\bibnamefont {Gupta}} \emph {et~al.},\ }\bibfield  {title} {\bibinfo {title} {{Characterizing gravitational wave detector networks: from A$^\sharp$ to cosmic explorer}},\ }\href {https://doi.org/10.1088/1361-6382/ad7b99} {\bibfield  {journal} {\bibinfo  {journal} {Class. Quant. Grav.}\ }\textbf {\bibinfo {volume} {41}},\ \bibinfo {pages} {245001} (\bibinfo {year} {2024})},\ \Eprint {https://arxiv.org/abs/2307.10421} {arXiv:2307.10421 [gr-qc]} \BibitemShut {NoStop}%
\bibitem [{\citenamefont {Borhanian}(2021)}]{Borhanian:2020ypi}%
  \BibitemOpen
  \bibfield  {author} {\bibinfo {author} {\bibfnamefont {S.}~\bibnamefont {Borhanian}},\ }\bibfield  {title} {\bibinfo {title} {{GWBENCH: a novel Fisher information package for gravitational-wave benchmarking}},\ }\href {https://doi.org/10.1088/1361-6382/ac1618} {\bibfield  {journal} {\bibinfo  {journal} {Class. Quant. Grav.}\ }\textbf {\bibinfo {volume} {38}},\ \bibinfo {pages} {175014} (\bibinfo {year} {2021})},\ \Eprint {https://arxiv.org/abs/2010.15202} {arXiv:2010.15202 [gr-qc]} \BibitemShut {NoStop}%
\bibitem [{\citenamefont {Khadkikar}\ \emph {et~al.}(2025)\citenamefont {Khadkikar}, \citenamefont {Gupta}, \citenamefont {Kashyap}, \citenamefont {Chandra}, \citenamefont {Gamba},\ and\ \citenamefont {Sathyaprakash}}]{Khadkikar:2025ith}%
  \BibitemOpen
  \bibfield  {author} {\bibinfo {author} {\bibfnamefont {S.}~\bibnamefont {Khadkikar}}, \bibinfo {author} {\bibfnamefont {I.}~\bibnamefont {Gupta}}, \bibinfo {author} {\bibfnamefont {R.}~\bibnamefont {Kashyap}}, \bibinfo {author} {\bibfnamefont {K.}~\bibnamefont {Chandra}}, \bibinfo {author} {\bibfnamefont {R.}~\bibnamefont {Gamba}},\ and\ \bibinfo {author} {\bibfnamefont {B.}~\bibnamefont {Sathyaprakash}},\ }\bibfield  {title} {\bibinfo {title} {{Cosmic Calipers: Precise and Accurate Neutron Star Radius Measurements with Next-Generation Gravitational Wave Detectors}},\ }\href@noop {} {\  (\bibinfo {year} {2025})},\ \Eprint {https://arxiv.org/abs/2502.03463} {arXiv:2502.03463 [astro-ph.HE]} \BibitemShut {NoStop}%
\bibitem [{\citenamefont {Lai}(1994)}]{Lai:1993di}%
  \BibitemOpen
  \bibfield  {author} {\bibinfo {author} {\bibfnamefont {D.}~\bibnamefont {Lai}},\ }\bibfield  {title} {\bibinfo {title} {{Resonant oscillations and tidal heating in coalescing binary neutron stars}},\ }\href {https://doi.org/10.1093/mnras/270.3.611} {\bibfield  {journal} {\bibinfo  {journal} {Mon. Not. Roy. Astron. Soc.}\ }\textbf {\bibinfo {volume} {270}},\ \bibinfo {pages} {611} (\bibinfo {year} {1994})},\ \Eprint {https://arxiv.org/abs/astro-ph/9404062} {arXiv:astro-ph/9404062} \BibitemShut {NoStop}%
\bibitem [{\citenamefont {Steinhoff}\ \emph {et~al.}(2016)\citenamefont {Steinhoff}, \citenamefont {Hinderer}, \citenamefont {Buonanno},\ and\ \citenamefont {Taracchini}}]{Steinhoff:2016rfi}%
  \BibitemOpen
  \bibfield  {author} {\bibinfo {author} {\bibfnamefont {J.}~\bibnamefont {Steinhoff}}, \bibinfo {author} {\bibfnamefont {T.}~\bibnamefont {Hinderer}}, \bibinfo {author} {\bibfnamefont {A.}~\bibnamefont {Buonanno}},\ and\ \bibinfo {author} {\bibfnamefont {A.}~\bibnamefont {Taracchini}},\ }\bibfield  {title} {\bibinfo {title} {{Dynamical Tides in General Relativity: Effective Action and Effective-One-Body Hamiltonian}},\ }\href {https://doi.org/10.1103/PhysRevD.94.104028} {\bibfield  {journal} {\bibinfo  {journal} {Phys. Rev. D}\ }\textbf {\bibinfo {volume} {94}},\ \bibinfo {pages} {104028} (\bibinfo {year} {2016})},\ \Eprint {https://arxiv.org/abs/1608.01907} {arXiv:1608.01907 [gr-qc]} \BibitemShut {NoStop}%
\bibitem [{\citenamefont {Gamba}\ \emph {et~al.}(2021)\citenamefont {Gamba}, \citenamefont {Breschi}, \citenamefont {Bernuzzi}, \citenamefont {Agathos},\ and\ \citenamefont {Nagar}}]{Gamba:2020wgg}%
  \BibitemOpen
  \bibfield  {author} {\bibinfo {author} {\bibfnamefont {R.}~\bibnamefont {Gamba}}, \bibinfo {author} {\bibfnamefont {M.}~\bibnamefont {Breschi}}, \bibinfo {author} {\bibfnamefont {S.}~\bibnamefont {Bernuzzi}}, \bibinfo {author} {\bibfnamefont {M.}~\bibnamefont {Agathos}},\ and\ \bibinfo {author} {\bibfnamefont {A.}~\bibnamefont {Nagar}},\ }\bibfield  {title} {\bibinfo {title} {{Waveform systematics in the gravitational-wave inference of tidal parameters and equation of state from binary neutron star signals}},\ }\href {https://doi.org/10.1103/PhysRevD.103.124015} {\bibfield  {journal} {\bibinfo  {journal} {Phys. Rev. D}\ }\textbf {\bibinfo {volume} {103}},\ \bibinfo {pages} {124015} (\bibinfo {year} {2021})},\ \Eprint {https://arxiv.org/abs/2009.08467} {arXiv:2009.08467 [gr-qc]} \BibitemShut {NoStop}%
\bibitem [{\citenamefont {Aalbers}\ \emph {et~al.}(2024)\citenamefont {Aalbers} \emph {et~al.}}]{LZ:2024zvo}%
  \BibitemOpen
  \bibfield  {author} {\bibinfo {author} {\bibfnamefont {J.}~\bibnamefont {Aalbers}} \emph {et~al.} (\bibinfo {collaboration} {LZ}),\ }\bibfield  {title} {\bibinfo {title} {{Dark Matter Search Results from 4.2 Tonne-Years of Exposure of the LUX-ZEPLIN (LZ) Experiment}},\ }\href@noop {} {\  (\bibinfo {year} {2024})},\ \Eprint {https://arxiv.org/abs/2410.17036} {arXiv:2410.17036 [hep-ex]} \BibitemShut {NoStop}%
\bibitem [{\citenamefont {{LZ Collaboration}}(2025)}]{hepdata.155182.v2}%
  \BibitemOpen
  \bibfield  {author} {\bibinfo {author} {\bibnamefont {{LZ Collaboration}}},\ }\href@noop {} {\bibinfo {title} {{Dark Matter Search Results from 4.2 Tonne-Years of Exposure of the LUX-ZEPLIN (LZ) Experiment (Version 2)}}},\ \bibinfo {howpublished} {{HEPData (collection)}} (\bibinfo {year} {2025}),\ \bibinfo {note} {\url{https://doi.org/10.17182/hepdata.155182.v2}}\BibitemShut {NoStop}%
\bibitem [{\citenamefont {Zevin}\ \emph {et~al.}(2022)\citenamefont {Zevin}, \citenamefont {Nugent}, \citenamefont {Adhikari}, \citenamefont {Fong}, \citenamefont {Holz},\ and\ \citenamefont {Kelley}}]{Zevin:2022dbo}%
  \BibitemOpen
  \bibfield  {author} {\bibinfo {author} {\bibfnamefont {M.}~\bibnamefont {Zevin}}, \bibinfo {author} {\bibfnamefont {A.~E.}\ \bibnamefont {Nugent}}, \bibinfo {author} {\bibfnamefont {S.}~\bibnamefont {Adhikari}}, \bibinfo {author} {\bibfnamefont {W.-f.}\ \bibnamefont {Fong}}, \bibinfo {author} {\bibfnamefont {D.~E.}\ \bibnamefont {Holz}},\ and\ \bibinfo {author} {\bibfnamefont {L.~Z.}\ \bibnamefont {Kelley}},\ }\bibfield  {title} {\bibinfo {title} {{Observational Inference on the Delay Time Distribution of Short Gamma-Ray Bursts}},\ }\href {https://doi.org/10.3847/2041-8213/ac91cd} {\bibfield  {journal} {\bibinfo  {journal} {Astrophys. J. Lett.}\ }\textbf {\bibinfo {volume} {940}},\ \bibinfo {pages} {L18} (\bibinfo {year} {2022})},\ \Eprint {https://arxiv.org/abs/2206.02814} {arXiv:2206.02814 [astro-ph.HE]} \BibitemShut {NoStop}%
\bibitem [{\citenamefont {Carr}\ and\ \citenamefont {Kuhnel}(2020)}]{Carr:2020xqk}%
  \BibitemOpen
  \bibfield  {author} {\bibinfo {author} {\bibfnamefont {B.}~\bibnamefont {Carr}}\ and\ \bibinfo {author} {\bibfnamefont {F.}~\bibnamefont {Kuhnel}},\ }\bibfield  {title} {\bibinfo {title} {{Primordial Black Holes as Dark Matter: Recent Developments}},\ }\href {https://doi.org/10.1146/annurev-nucl-050520-125911} {\bibfield  {journal} {\bibinfo  {journal} {Ann. Rev. Nucl. Part. Sci.}\ }\textbf {\bibinfo {volume} {70}},\ \bibinfo {pages} {355} (\bibinfo {year} {2020})},\ \Eprint {https://arxiv.org/abs/2006.02838} {arXiv:2006.02838 [astro-ph.CO]} \BibitemShut {NoStop}%
\bibitem [{\citenamefont {Calder{\'o}n~Bustillo}\ \emph {et~al.}(2025)\citenamefont {Calder{\'o}n~Bustillo}, \citenamefont {del Rio}, \citenamefont {Sanchis-Gual}, \citenamefont {Chandra},\ and\ \citenamefont {Leong}}]{CalderonBustillo:2024akj}%
  \BibitemOpen
  \bibfield  {author} {\bibinfo {author} {\bibfnamefont {J.}~\bibnamefont {Calder{\'o}n~Bustillo}}, \bibinfo {author} {\bibfnamefont {A.}~\bibnamefont {del Rio}}, \bibinfo {author} {\bibfnamefont {N.}~\bibnamefont {Sanchis-Gual}}, \bibinfo {author} {\bibfnamefont {K.}~\bibnamefont {Chandra}},\ and\ \bibinfo {author} {\bibfnamefont {S.~H.~W.}\ \bibnamefont {Leong}},\ }\bibfield  {title} {\bibinfo {title} {{Testing Mirror Symmetry in the Universe with LIGO-Virgo Black-Hole Mergers}},\ }\href {https://doi.org/10.1103/PhysRevLett.134.031402} {\bibfield  {journal} {\bibinfo  {journal} {Phys. Rev. Lett.}\ }\textbf {\bibinfo {volume} {134}},\ \bibinfo {pages} {031402} (\bibinfo {year} {2025})},\ \Eprint {https://arxiv.org/abs/2402.09861} {arXiv:2402.09861 [gr-qc]} \BibitemShut {NoStop}%
\end{thebibliography}%
\end{document}